\documentclass{aa}
\usepackage{subfigure}
\usepackage{supertabular}
\usepackage{lscape}
\usepackage{bm}
\bibpunct{(}{)}{;}{a}{}{,} 
\usepackage{soul}
\setstcolor{yellow}
\usepackage{bm}
\usepackage{graphicx}
\usepackage{txfonts}
\def\pa{Pad$\rm \acute{e}_{(2,1)}$ }

\begin{document} 
	
	\label{firstpage}

	\title{Testing cosmic anisotropy with Pad$\rm \acute{e}$ approximation and Pantheon+ sample }
	\titlerunning{Cosmological applications}
	
	\author{J. P. Hu\inst{1} 
		\and J. Hu\inst{2} 
		\and X. D. Jia\inst{1}
		\and B. Q. Gao\inst{3} 
		\and F. Y. Wang\inst{1,4} 
	}
	
	\institute{School of Astronomy and Space Science, Nanjing University, Nanjing 210093, China\\
		\email{fayinwang@nju.edu.cn}
		\and
		School of Engineering, Dali University, Dali 671003, China 
		\and 
		Research Center for Astronomical Computing, Zhejiang Lab, Hangzhou, 311100, China
		\and 
		Key Laboratory of Modern Astronomy and Astrophysics (Nanjing University), Ministry of Education, Nanjing 210093, China
		}
	
	\date{Received date; accepted date}
	
	
\abstract
{Cosmography can be used to constrain the kinematics of universe in a model-independent way. In this work, we attempted to combine the Pad$\rm \acute{e}$ approximations with the latest Pantheon+ sample for testing cosmological principle. Based on the Pad$\rm \acute{e}$ approximations, we first gave the cosmographic constraints on the different order polynomials including third-order (Pad$\rm \acute{e}$$_{(2,1)}$), fourth-order (Pad$\rm \acute{e}$$_{(2,2)}$) and fifth-order (Pad$\rm \acute{e}$$_{(3,2)}$). The statistical analyses show that the Pad$\rm \acute{e}$$_{(2,1)}$ polynomial has the best performance. Its best-fits are $H_{0}$ = 72.53$\pm$0.28 km s$^{-1}$ Mpc$^{-1}$, $q_{0}$ = $-$0.35$_{-0.07}^{+0.08}$, and $j_{0}$ = 0.43$_{-0.56}^{+0.38}$. By further fixing $j_{0}$ = 1.00, it can be found that the Pad$\rm \acute{e}$$_{(2,1)}$ polynomial can describe the Pantheon+ sample better than the regular Pad$\rm \acute{e}$$_{(2,1)}$ polynomial and the usual cosmological models (including $\Lambda$CDM model, $w$CDM model, CPL model and $R_h$ = ct model). Based on the Pad$\rm \acute{e}$$_{(2,1)}$ ($j_{0}$ = 1) polynomial and hemisphere comparison (HC) method, we tested the cosmological principle and found the preferred directions of cosmic anisotropy, such as (l, b) = (304.6$^{\circ}$$_{-37.4}^{+51.4}$, $-$18.7$^{\circ}$$_{-20.3}^{+14.7}$) and (311.1$^{\circ}$$_{-8.4}^{+17.4}$, $-$17.53$^{\circ}$$_{-7.7}^{+7.8}$) for $q_{0}$ and $H_{0}$, respectively. These two directions are consistent with each other in $1\sigma$ confidence level, but the corresponding results of statistical isotropy analyses including Isotropy and Isotropy with real positions (RP) are quite different. The statistical significance of $H_{0}$ are stronger than that of $q_{0}$, i.e., 4.75$\sigma$ and 4.39$\sigma$ for the Isotropy and Isotropy with RP respectively. Reanalysis with fixed $q_{0} = -0.55$ (corresponds to $\Omega_{m}$ = 0.30) gives similar results. Overall, our model-independent results provide clear indications for a possible cosmic anisotropy, which must be taken seriously. Further test is needed to better understand this signal.}
	
\maketitle
			
%
\section{Introduction} \label{sec:intro} 
Cosmography has been widely used in cosmological data processing to restrict the state of the kinematics of our Universe in a model-independent way \citep{2015CQGra..32m5007V,2016IJGMM..1330002D,2018MNRAS.476.3924C,2019MNRAS.484.4484C,2019IJMPD..2830016C,2019A&A...628L...4L,2021A&A...649A..65B}. It relies only on the assumption of a homogeneous and isotropy universe as described by the Friedman-Lemaitre-Robertson-Walker (FLRW) metric \citep{1972gcpa.book.....W}. Its methodology is essentially based on expanding measurable cosmological quantity into Taylor series around the present time. In this framework, the evolution of universe can be described by some cosmographic parameters, such as Hubble parameter $H$, deceleration $q$, jerk $j$, snap $s$, and lerk $l$ parameters. The corresponding definitions of them can be expressed as follows:
	\begin{eqnarray}
		\label{eq:q0j0}
		H = \frac{\dot{a}}{a}, q = -\frac{1}{H^{2}}\frac{\ddot{a}}{a}, j=\frac{1}{H^3}\frac{\dot{\ddot{a}}}{a},
		s=\frac{1}{H^4}\frac{\ddot{\ddot{a}}}{a}, l=\frac{1}{H^5}\frac{\dot{\ddot{\ddot{a}}}}{a}\label{j}.
	\end{eqnarray} 
	
Various methods have been proposed in the literature. The first proposed method is $z$-redshift \citep{2004JCAP...09..009C,2004ApJ...607..665R,2004CQGra..21.2603V}, which estimated the cosmic evolution as $z \sim 0$ well, but it failed at high redshifts \citep{1998tx19.confE.276C,2004ApJ...607..665R,2004CQGra..21.2603V,Wang2009,2010JCAP...03..005V}. The main reason is that the data are far from the limits of Taylor expansions. In cosmography, it is called the convergence problem. Focused on this issue, many improved methods were proposed to solve this problem, such as $y$-redshift \citep{2001IJMPD..10..213C,2003PhRvL..90i1301L,2010JCAP...03..005V}, $E(y)$ \citep{2020ApJ...900...70R}, $\log{(1+z)}$ \citep{2019NatAs...3..272R,2020PhRvD.102l3532Y}, $\log{(1+z)}+k_{ij}$ \citep{2021A&A...649A..65B}, Pad$\rm \acute{e}$ approximations \citep{2014PhRvD..89j3506G,2014JCAP...01..045W,2020MNRAS.494.2576C} and Chebyshev approximations \citep{2018MNRAS.476.3924C}. In theory, the improved methods can effectively avoid the convergence problem. Combined analyses with the cosmic observations show that things are not simple \citep{2017EPJC...77..434Z,2020AA...643A..93H,2020ApJ...900...70R,2022A&A...661A..71H}. \citet{2017EPJC...77..434Z} found that the $y$-redshift produces larger variances beyond the second-order expansion from the analyses of the Joint Light-curve Analysis sample \citep[JLA;][]{2014A&A...568A..22B}. Similar result was also found by \citet{2020ApJ...900...70R} using a composite sample which including the type Ia supernovae (SNe Ia), quasars and gamma-ray bursts (GRBs). They attempted to recast $E(z)$ as a function of $y= z/(1+z)$ and adopted the new series expansion of the $E(y)$ function to compare dark energy models. For the convergence problem of the cosmographic method, there are still many details need to be further studied.
	
For the different cosmographic techniques, there has been many works to compare their superiority by utilizing different observational data \citep{2018MNRAS.476.3924C,2020MNRAS.494.2576C,2022A&A...661A..71H,2022PhRvD.106l3523P}. Utilizing the Akaike information criterion \citep[AIC;][]{1100705} and the Bayesian information criterion \citep[BIC;][]{10.1214/aos/1176344136}, \citet{2020MNRAS.494.2576C} critically compared the auxiliary variables with the rational approximations, including $y_1 = 1 - a$, $y_2 = \arctan(a^{-1} -1 )$ and the Pad$\rm \acute{e}$ approximations. They found that even though $y_2$ overcomes the issues of $y_1$, the performance of Pad$\rm \acute{e}$ approximations is better than that of the auxiliary variables $y_1 = 1 - a$ and $y_2 = \arctan(a^{-1} -1 )$. Meanwhile, they also made Monte Carlo analyses combining the Pantheon sample, $H(z)$ and shift parameter measurements, and concluded that the \pa polynomials is statistically the optimal approach to explain low and high-redshift observations. Utilizing the Pantheon sample \citep{2018ApJ...859..101S} and 31 long gamma-ray bursts \citep{2022ApJ...924...97W}, \citet{2022A&A...661A..71H} made a systematic comparison among the commonly used cosmographic expansions (including $z$-redshift, $y$-redshift, $E(y)$, $\log(1+z)$, $\log(1+z)+k_{ij}$ and Pad$\rm \acute{e}$ approximations) with different expansion orders. They found that the expansion order can significantly affect the results, especially the $y$-redshift method. \pa polynomial is suitable for both low and high redshift cases. Pad$\rm \acute{e}_{(2,2)}$ polynomial performs well in high redshift case. According to the statistical results, they concluded that it is important to choose the suitable expansion method and order based on the sample used. From the previous comparative researches, it can be found that the Pad$\rm \acute{e}$ approximations have better performance than the traditional Taylor series. 
	
	\begin{figure*}[htp]
		\centering
		\includegraphics[width=0.35\textwidth]{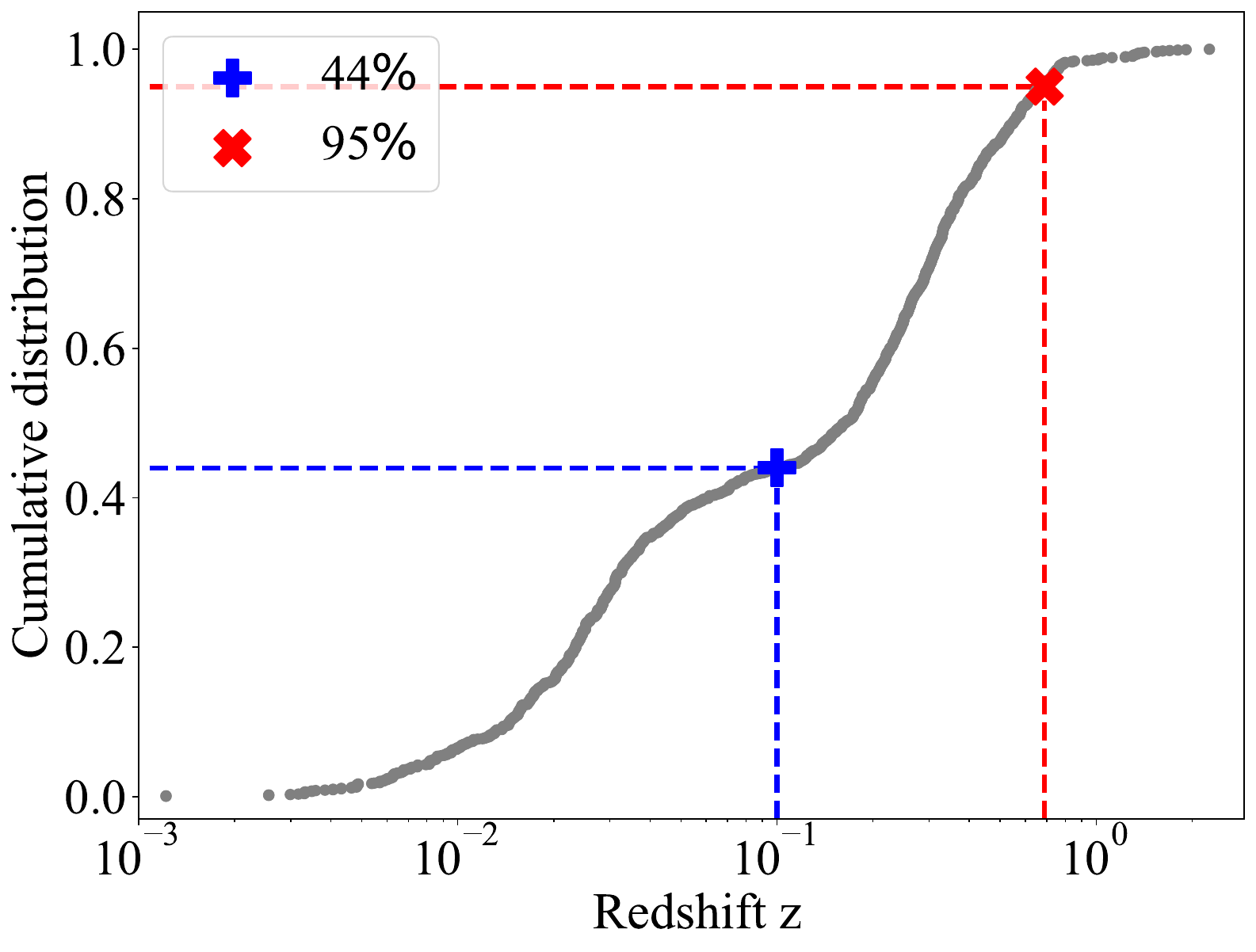} \
		\includegraphics[width=0.5\textwidth]{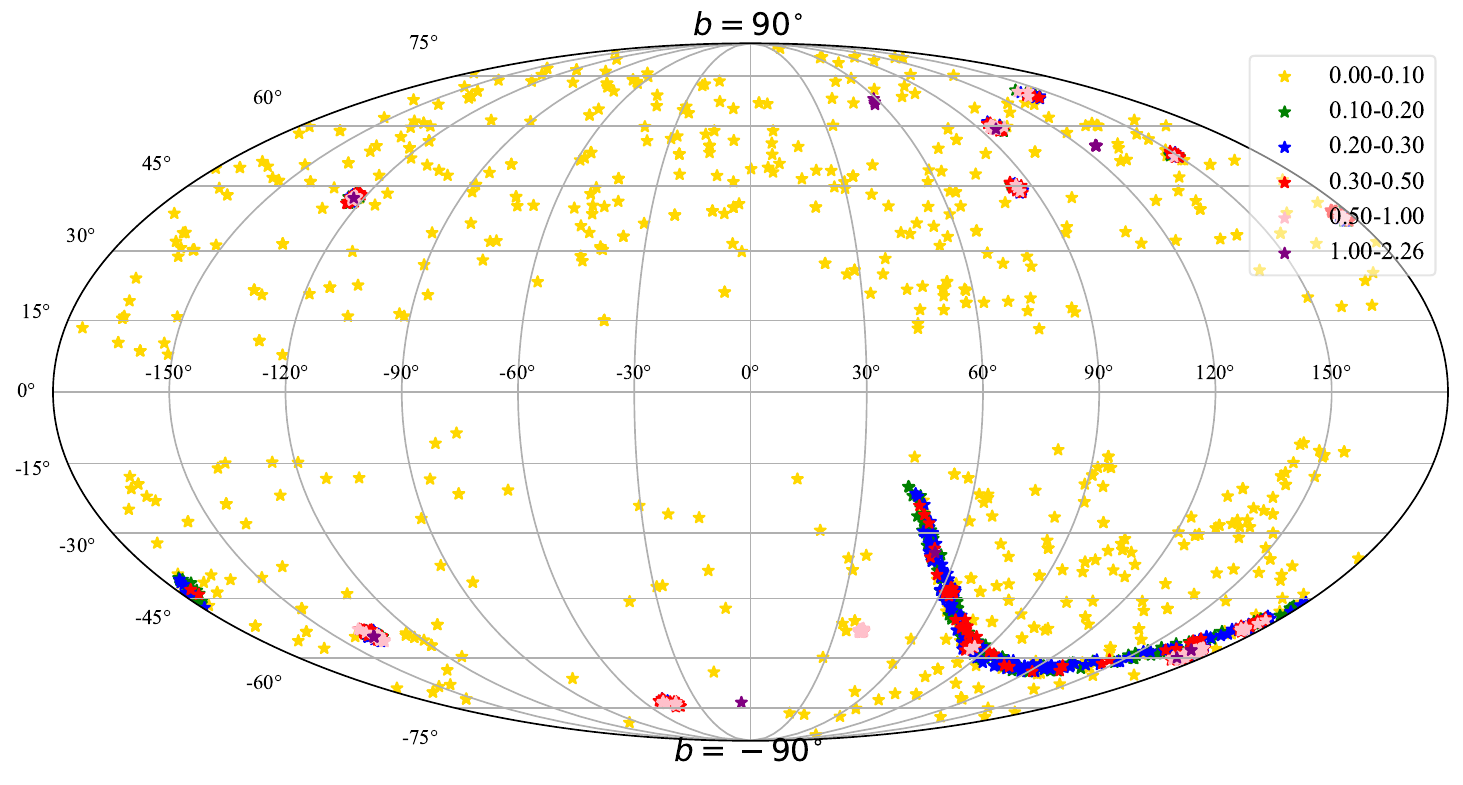}
		\caption{Basic information about Pantheon+ sample. The left panel shows the cumulative redshift distribution. The right panel shows the location of different redshift SNe in the galactic coordinate system.}
		\label{F1}       
	\end{figure*} 
	
Recently, the latest sample of SNe Ia (Pantheon+) is released by the SH0ES collaboration \citep{2022ApJ...938..110B,2022ApJ...938..113S}. It is the updated version of Pantheon sample, and covers redshift range (0.001, 2.26). This latest sample has been widely used for cosmological applications, such as cosmological constraints \citep{2023PhRvD.107j3521C,2023ApJ...951...63D,2023MNRAS.523.3406F,2023PDU....4001224P,2023PhRvD.108f3522Q,2023ApJ...954L..31S}, non-Gaussian likelihoods \citep{2023arXiv230306974D,2023arXiv231202075L}, absolute magnitudes \citep{2024ApJ...964L...4C,2023MNRAS.520.5110P}, modified gravity \citep{2023JCAP...06..039D,2023PDU....4201281K,2023EPJC...83..442V}, dynamical dark energy \citep{2023arXiv231116862T,2023arXiv230504946V}, running vacuum in Universe \citep{2023Univ....9..262S,2023Univ....9..204Y,2024CQGra..41a5026G}, Hubble tension \citep{2023A&A...674A..45J,2023arXiv231016800S,2023PDU....4201365Y,2024PhRvD.109b3527A}, $\sigma_8$ tension \citep{2023PhRvD.107l3538P}, cosmological principle \citep{2023PhRvD.108l3533M,2023PhRvD.108f3509P,2023ChPhC..47l5101T,2024arXiv240217741B,2024AA...681A..88H} and so on. The cosmological principle is the basic assumption of the cosmology, which requires that the Universe is statistically isotropic and uniform on a sufficiently large scale. But there has been many researches suggest that the Universe may be inhomogeneous and anisotropic, including the fine-structure constant \citep{2023MmSAI..94b.270M}, the direct measurement of the Hubble parameter \citep{2021PhRvL.126w1101K}, and source counts \citep{2023MNRAS.524.3636S}. Recent year, SNe Ia have been widely employed to test the cosmological principle \citep{2018MNRAS.478.5153S,2019A&A...631L..13C,2023PhRvD.107b3507K,2023PhRvD.108l3533M,2023PhRvD.108f3509P,2023ChPhC..47l5101T,2024AA...681A..88H}. \citet{2023ChPhC..47l5101T} fitted the full Pantheon+ data with the dipole-modulated $\Lambda$CDM model and found that the dipole appears at the 2$\sigma$ confidence level if $z_{c} \leq 0.1$. \citet{2023PhRvD.108f3509P} used the hemisphere comparison (HC) method to test the isotropy of the SNe Ia absolute magnitudes of the Pantheon+ and SH0ES samples in various redshift/distance bins. They found sharp changes of the level of anisotropy occurring at distances less than 40 Mpc. Meanwhile, \citet{2023PhRvD.108l3533M} also employed the Pantheon+ sample to analyse the anisotropic distance ladder, and found a larger $H_{0}$ near by the CMB dipole direction. They pointed that the cosmic anisotrpy may be due to a breakdown in the cosmological principle \citep{2022PhRvD.105f3514K}, or mundanely due to a statistical fluctuation in a small sample of SN in Cepheid host galaxies. Employing on the Pantheon+ sample, \citet{2024AA...681A..88H} proposed a new method which named the region-fitting method to map the all-sky distribution of cosmological parameters to examine the cosmological principle. They found a local matter underdensity region (${308.4^{\circ}}$$_{-48.7}^{+47.6}$, ${-18.2^{\circ}}$$_{-28.8}^{+21.1}$) and a preferred direction of the cosmic anisotropy (${313.4^{\circ}}$$_{-18.2}^{+19.6}$, ${-16.8^{\circ}}$$_{-10.7}^{+11.1}$) in galactic coordinates. Through investigating the correlation between the degree of cosmic anisotropy and the redshift, they also proposed that the $H_{0}$ evolution \citep{2020PhRvD.102j3525K,2020A&A...639A.101M,2020MNRAS.498.1420W,2021ApJ...912..150D,2022MNRAS.517..576H,2023A&A...674A..45J,2024EPJC...84..317M} might be associated with the violation of the cosmological principle. At present, the anisotropy of the accelerating expansion of the universe is still unresolved \citep{2021AA...649A.151M,2022JPhCS2191a2001A,2022JHEAp..34...49A,2022A&A...668A..34H,2022NewAR..9501659P,2023JCAP...10..050B,2023CQGra..40i4001K,2024arXiv240112291K,2024Astro...3...43N,2024arXiv240216885Y}. Most of above researches of the cosmological principle are based on the $\Lambda$CDM model or other extended model. Until now, no research has considered using the Pad$\rm \acute{e}$ approximation to probing cosmological principle.

	In this paper, we utilized the latest Pantheon+ sample and the Pad$\rm \acute{e}$ approximation for cosmological applications including cosmographic constraints, model comparison and cosmological principle. Firstly, we gave the cosmographic constraints, and try to find out the suitable expansion order for the Pantheon+ sample. After that, we critically compared the Pad$\rm \acute{e}$ approximation with common cosmological models, such as flat $\Lambda$CDM model, non-flat $\Lambda$CDM model, $w$CDM model, Chevallies-Polarski-Linder (CPL) model \citep{2001IJMPD..10..213C,2003PhRvL..90i1301L}, and $R_h$ = ct model \citep{2012MNRAS.419.2579M}. In addition, we mapped the anisotropy level in the galactic coordinate system utilizing the Pantheon+ sample based on the Pad$\rm \acute{e}$ approximation and the HC method. For strict consideration, the corresponding statistical isotropy analyses were done. The outline of this paper is as follows. In section \ref{data}, we describe the observational data and fitting method. Theoretical models used in this article are introduced in section \ref{models}. In section \ref{method}, we describe the hemispheric comparison method used to probe the preferred direction of cosmic anisotropy.  Afterwards, we present and discuss the main results in section \ref{result}. Finally, a brief summary is given in section \ref{conl}.

	\section{Observational data and fitting method} \label{data}
	\subsection{Pantheon+ sample}
	Pantheon+ sample is the latest SNe Ia dataset \citep{2022ApJ...938..110B,2022ApJ...938..113S}. It consists of 1701 SNe Ia light curves observed from 1550 distinct, and covers redshift range (0.001, 2.26). In Fig. \ref{F1}, we gave some basic information including the cumulative distribution of redshift (left panel) and the location distribution in galactic coordinate system (right panel). From the left panel of Fig. \ref{F1}, it is easy to find that the redshift of nearly half observations (44\%) are below 0.10, and the most of observations (95\%) are below 0.70. From the right panel of Fig. \ref{F1}, we can find that the belt-like structure still plays the most important role in the Pantheon+ sample although the number of new added SNe is nearby 700. Combining the redshift information, it can be found that the distribution of SNe whose redshift smaller than 0.10 is relatively homogeneous. However, the distribution of higher redshift SNe is relatively concentrated, because they are mainly obtained from some high redshift surveys, such as the Sloan Digital Sky Survey \citep[SDSS;][]{2018PASP..130f4002S}, the Panoramic Survey Telescope \& Rapid Response System Medium Deep Survey \citep[PS1MD;][]{2018ApJ...859..101S}, the SuperNova Legacy Survey \citep[SNLS;][]{2014A&A...568A..22B}, and so on. In a nutshell, the SNe distribution of Pantheon+ sample is still inhomogeneous. In addition, it is note that SNe Ia have a nearly uniform intrinsic luminosity with an absolute magnitude around $M$ $\sim$ $-$19.5 \citep{2001LRR.....4....1C} which make it to a well-established class of standard candles. The difficulty in the cosmological applications lies in the identification of absolute magnitude $M$, due to different sources of systematic and statistical uncertainties. In this paper, the systematic ($\mathbf{C}_\mathrm{sys}$) and statistical ($\mathbf{C}_\mathrm{stat}$) covariance matrices are considered. The used datasets, distance modulus $\mu_{obs}$ and total covariance matric $\mathbf{C}_\mathrm{stat+sys}$ are provided by \citet{2022ApJ...938..110B} and publicly available\footnote{https://github.com/PantheonPlusSH0ES/DataRelease}. The observational distance modulus is calibrated by the second rung of the distance ladder using Cepheids measuring the SNe Ia absolute magnitude as $M_B$ = -19.25 $\pm$ 0.01 \citep{2022ApJ...938..110B,2022ApJ...934L...7R,2024arXiv240507039B}.

	\subsection{Fitting method}
	The best fits of the cosmological parameters are derived by minimizing the chi-square ($\chi^{2}$),
	\begin{equation}
		\chi^{2} = (\mu_\mathrm{obs} - \mu_\mathrm{th}) \, \mathbf{C}^{-1}_\mathrm{stat+syst} \, (\mu_\mathrm{obs} - \mu_\mathrm{th}) ^\mathrm{T} ,
		\label{chi}
	\end{equation}
	where $\mu_\mathrm{obs}$ is the observational distance modulus, and $\mu_\mathrm{th}$ is the theoretical distance modulus which can be obtained from following formula:  
	\begin{equation}
		\mu_\mathrm{th}(z_{i}, P_{i}) = m - M = 5 \log_{10} \frac{d_{\rm L}(z_{i}, P_{i})}{\textnormal{Mpc}} + 25.
		\label{muth}
	\end{equation}
	Here, $z_{i}$ is the peculiar-velocity-corrected CMB-frame redshift of each SN \citep{2022PASA...39...46C}, $P_{i}$ represents the parameters to be fitted, $M$ is the absolute magnitude, $m$ is the apparent magnitude of the source, and $d_{\rm L}$ is the luminosity distance. For the calculation of $d_{\rm L}$, we need to fix a cosmological model or expansion method. In here, we briefly introduce how to obtain $d_{L}$ based on the Pad$\rm \acute{e}$ approximations and some usual cosmological models, including Pad$\rm \acute{e}$$_{(2,1)}$, Pad$\rm \acute{e}$$_{(2,2)}$, Pad$\rm \acute{e}$$_{(3,2)}$, flat $\Lambda$CDM model, non-flat $\Lambda$CDM model, $w$CDM model, CPL model, and $R_{\rm h}$ = ct model.

	\section{Theoretical models} \label{models}
	\textbf{Pad$\rm \acute{e}$ approximations.} The Pad$\rm \acute{e}$ approximations \citep{1992A} was built up from the standard Taylor definition and can be used to lower divergences at $z \geq$ 1. It often gives a better approximation for the function than truncating its Taylor series, and it may still work where the Taylor series does not converge \citep{2014JCAP...01..045W}. Due to its excellent convergence properties, the Pad$\rm \acute{e}$ polynomials have been considered at high redshifts in cosmography \citep{2014PhRvD..89j3506G,2017A&A...598A.113D,2018JCAP...05..008C,2020MNRAS.494.2576C}. The Taylor expansion of a generic function $f(z)$ can be described by a given function $f(z) = \int_{i=0}^{\infty} c_{i}z^{i}$, where $c_{i} = f^{i}(0)/i!$, which is approximated by means of a $(n, m)$ Pad$\rm \acute{e}$ approximation by the radio polynomial \citep{2020MNRAS.494.2576C}  
	\begin{eqnarray}
		\label{eq:Pnm}
		P_{n, m}(z) = \frac{\sum_{i = 0}^{n} a_{i}z^{i}}{1 + \sum_{j = 1}^{m} b_{j}z^{j}};
	\end{eqnarray} 
	there are a total $(n + m + 1)$ number of independent coefficients. In the numerator, we have $n + 1$ independent coefficients, whereas in the denominator there is $m$. Since, by construction, $b_{0} = 1$ is required, we have
	\begin{eqnarray}
		\label{eq:ff}
		f(z) -P_{n , m} (z) = \textit{O}(z^{n+m+1}). 
	\end{eqnarray}
	The coefficients $b_{i}$ in Eq. (\ref{eq:Pnm}) were thus determined by solving the follow homogeneous system of linear equations \citep{1993Litvinov}:
	\begin{eqnarray}
		\label{eq:bb0}
		\sum_{j=1}^{m} b_{j}c_{n+k+j} = -b_{0}c_{n+k},
	\end{eqnarray} 
	which is valid for $k = 1,...,m$. All coefficients $a_{i}$ in Eq. (\ref{eq:Pnm}) can be computed using the formula
	\begin{eqnarray}
		\label{eq:ai}
		a_{i} = \sum_{k=0}^{i} b_{i-k}c_{n+k}. 
	\end{eqnarray} 
	In terms of the investigations of \citet{2020MNRAS.494.2576C} on the Pad$\rm \acute{e}$ approximation, we finally chose to use the Pad$\rm \acute{e}$$_{(2,1)}$, Pad$\rm \acute{e}$$_{(2,2)}$ and Pad$\rm \acute{e}$$_{(3,2)}$ polynomials to represent the third-order, the fourth-order and the fifth-order polynomials, respectively. More detailed information about the selections of the specific polynomials can be found in \citet{2020MNRAS.494.2576C}. The corresponding luminosity distances are given as follows. \\
	(1) Pad$\rm \acute{e}$$_{(2,1)}$ polynomial of the luminosity distance: 
	\begin{eqnarray}
		\label{eq:dlp21}
		d_{\rm L} = \frac{c}{H_{0}}[ \frac{z(6(-1+q_{0})+(-5-2j_{0}+q_{0}(8+3q_{0}))z)}{-2(3+z+j_{0}z)+2q_{0}(3+z+3q_{0}z)}].
	\end{eqnarray}
	(2) Pad$\rm \acute{e}$$_{(2,2)}$ polynomial of the luminosity distance: 
	\begin{eqnarray}
		\label{eq:dlp22}
		d_{\rm L} &=& \frac{c}{H_0}[6z(10 + 9 z - 6 q_0^3 z + s_0 z - 2 q_0^2 (3 + 7 z) \nonumber \\
		&-& q_0 (16 + 19 z) +
		j_0 (4 + (9 + 6 q_0) z))\Big/(60 + 24 z \nonumber \\
		&+& 6 s_0 z - 2 z^2     + 4 j_0^2 z^2 - 9 q_0^4 z^2 - 3 s_0 z^2 \nonumber \\
		&+& 6 q_0^3 z (-9 + 4 z) + q_0^2 (-36 - 114 z + 19 z^2)  \nonumber \\
		&+&j_0 (24 + 6 (7 + 8 q_0) z + (-7 - 23 q_0 + 6 q_0^2) z^2) \nonumber \\
		&+&  q_0 (-96 - 36 z + (4 + 3 s_0) z^2))].
	\end{eqnarray}
	(3) Pad$\rm \acute{e}$$_{(3,2)}$ polynomial of the luminosity distance: 
	\begin{eqnarray}
		\label{eq:dlp32}
		d_{\rm L} &=&\frac{c}{H_0}[z (-120 - 180 s_0 - 156 z - 36 l_0 z - 426 s_0 z \nonumber \\
		&-& 40 z^2 + 80 j_0^3 z^2 - 30 l_0 z^2 - 135 q_0^6 z^2	- 210 s_0 z^2 \nonumber \\
		&+& 15 s_0^2 z^2 - 270 q_0^5 z (3 + 4 z) + 9 q_0^4 (-60 + 50 z + 63 z^2) \nonumber \\ 
		&+& 2 q_0^3 (720 + 1767 z	+ 887 z^2) + 3 j_0^2 (80 + 20 (13 + 2 q_0) z \nonumber \\
		&+& (177 + 40 q_0 - 60 q_0^2) z^2) + 6 q_0^2 (190 + 5 (67 + 9 s_0) z \nonumber \\
		&+& (125 + 3 l_0 + 58 s_0) z^2) -6 q_0 (s_0 (-30 + 4 z + 17 z^2) \nonumber \\
		&-& 2 (20 + (31 + 3 l_0) z + (9 + 4 l_0) z^2)) + 6 j_0 (-70 \nonumber \\
		&+& (-127 + 10 s_0) z + 45 q_0^4 z^2 + (-47 - 2 l_0 + 13 s_0) z^2 \nonumber \\
		&+& 5 q_0^3 z (30 + 41 z) - 3 q_0^2 (-20 + 75 z + 69 z^2) \nonumber \\
		&+& 2 q_0 (-115 - 274 z + (-136 + 5 s_0) z^2)))]\Big/[3(-40 \nonumber \\
		&-& 60 s_0 - 32 z - 12 l_0 z - 112 s_0 z - 4 z^2 + 40 j_0^3 z^2 - 4 l_0 z^2 \nonumber \\
		&-& 135 q_0^6 z^2 - 24 s_0 z^2 + 5 s_0^2 z^2 - 30 q_0^5 z (12 + 5 z) \nonumber \\
		&+& 3 q_0^4 (-60 + 160 z + 71 z^2) +j_0^2 (80 + 20 (11 + 4 q_0) z \nonumber \\
		&+& (57 + 20 q_0 - 40 q_0^2) z^2) + 6 q_0^3 (80 + 188 z + (44 \nonumber \\
		&+& 5 s_0) z^2) + 2 q_0^2 (190 + 20 (13 + 3 s_0) z + (46 + 6 l_0 \nonumber \\
		&+& 21 s_0) z^2)+4 q_0 (20 + (16 + 3 l_0) z + (2 + l_0) z^2 \nonumber \\
		&+& s_0 (15 - 17 z - 9 z^2))+2 j_0 (-70 + 2 (-46 + 5 s_0) z \nonumber \\
		&+& 90 q_0^4 z^2 + (-16 - 2 l_0 + 3 s_0) z^2 + 15 q_0^3 z (12 + 5 z) \nonumber \\
		&+& q_0^2 (60 - 370 z - 141 z^2) + 2 q_0 (-115 - 234 z \nonumber \\
		&+& 2 (-26 + 5 s_0) z^2)))],
	\end{eqnarray}
	where $c$ is the speed of light, $H_{0}$, $q_{0}$, $j_{0}$, $s_{0}$ and $l_{0}$ represent current values. 
	
	\textbf{Flat $\Lambda$CDM model}. Considering the flat $\Lambda$CDM model, the corresponding luminosity distance $d_{\rm L}$ can be calculated from \citep{2021MNRAS.507..730H}
	\begin{equation}
		d_{\rm L} = \frac{c(1+z)}{H_{0}} \int_{0}^{z} 
		\frac{dz'}{\sqrt{\Omega_{m} (1+z')^{3} + \Omega_{\Lambda}}},
		\label{flatL}
	\end{equation}
	where $\Omega_{m}$ is the matter density, and $\Omega_{\Lambda}$ is the dark energy density. 
	
	\textbf{Non-flat $\Lambda$CDM model.} For the nonflat $\Lambda$CDM model, the corresponding luminosity distance $d_{\rm L}$ should be written as \citep{2021MNRAS.507..730H}
	\begin{eqnarray}
		\label{nonflat}
		d_{\rm L} =
		\begin{cases}
			\frac{c(1+z)}{H_{0}}(-\Omega_{\rm k})^{-\frac{1}{2}} \sin{[(-\Omega_{\rm k})^{\frac{1}{2}} \int_{0}^{z} \frac{dz'}{E(z')}]}, &\Omega_{\rm k} < 0, \\
			\frac{c(1+z)}{H_{0}} \int_{0}^{z} \frac{dz'}{E(z')}, &\Omega_{\rm k} = 0,\\
			\frac{c(1+z)}{H_{0}}\Omega_{\rm k}^{-\frac{1}{2}} \sinh{[\Omega_{\rm k}^{\frac{1}{2}} \int_{0}^{z} \frac{dz'}{E(z')}]}, &\Omega_{\rm k} > 0,
		\end{cases}
	\end{eqnarray}
	\begin{eqnarray}
		\label{eq:ez}
		E(z') = \sqrt{\Omega_{m} (1+z')^{3} + (1-\Omega_{m}-\Omega_{\Lambda})(1+z')^{2} + \Omega_{\Lambda}},
	\end{eqnarray}
	here, $\Omega_{\rm k}$ is the spatial curvature.

	\textbf{$w$CDM model.} For the $w$CDM model, the corresponding luminosity distance $d_{\rm L}$ can be written as \citep{2023MNRAS.521.4406L}
	\begin{equation}
		d_{\rm L} = \frac{c(1+z)}{H_{0}} \int_{0}^{z} 
		\frac{dz'}{\sqrt{\Omega_{m} (1+z')^{3} + \Omega_{\Lambda} (1+z)^{3(1+w_{0})}}},
		\label{wcdm}
	\end{equation}
	where, $w_{0}$ represents the constant Eqution of State (Eos). When $w_{0} \simeq -1$, this model regresses to the flat $\Lambda$CDM model.

	\textbf{CPL model.} Rewrite Eq. (\ref{wcdm}), we can obtain the corresponding expression of luminosity distance for CPL model \citep{2001IJMPD..10..213C,2003PhRvL..90i1301L}
	\begin{equation}
		d_{\rm L} = \frac{c(1+z)}{H_{0}} \int_{0}^{z} 
		\frac{dz'}{\sqrt{\Omega_{m} (1+z')^{3} + \Omega_{\Lambda} (1+z)^{3(1+w_{0}+w_{a})} \times
				 \exp^{-\frac{3w_{a}z}{1+z}}}}.
		\label{wacdm}
	\end{equation}
	In the CPL model, dark energy evolves with redshift as parameterized Eos, $w=w_{0} + w_{a}z/(1+z)$.
	
	$\bm{R_{h}}$\textbf{ = ct model.} For the $R_{h}$ = ct model \citep{2012MNRAS.419.2579M}, the corresponding luminosity distance is as follow:
	\begin{equation}
		d_{\rm L} = R_{h}(t_{0})(1+z)\ln{(1+z)},
		\label{rhct}
	\end{equation}
	where, $R_{h}(t_{0})$ is the current gravitational horizon, which can also be represented by $c/H_{0}$. 
	
	Combining Eqs. (\ref{chi}), (\ref{muth}) and the expressions of luminosity distance, we can give the corresponding constraints for the different expansion methods or cosmological models. The comparison results of different constraints are obtained by utilizing the Akaike information criterion \citep[AIC;][]{10.1214/aos/1176344136} and the Bayesian information criterion \citep[BIC;][]{1100705}, which are the last set of techniques that can be employed for model comparison based on information theory. The corresponding definitions read as follows:  
	\begin{eqnarray}
		AIC &=& 2n - 2\ln{\mathcal{L}} ,\\
		BIC &=& n\log{N} - 2\ln{\mathcal{L}} ,
		\label{eq:AIC}
	\end{eqnarray}
	where $n$ is the number of free parameters, $N$ is the total number of data points, and $\mathcal{L}$ is the maximum value of the likelihood function. The values of $\mathcal{L}$ are derived from the follow formula,
	\begin{eqnarray}
		\mathcal{L} = \exp ({-0.5\chi_{min}^{2}}),
		\label{eq:l}
	\end{eqnarray}
	where, $\chi_{min}^{2}$ is the value of $\chi^{2}$ calculated with the best fitting results. The model that has lower values of AIC and BIC will be the suitable model for the employed data-set. Moreover, we also calculated the differences between $\Delta$AIC and $\Delta$BIC with respect to the corresponding flat $\Lambda$CDM values to measure the amount of information lost by adding extra parameters in the statistical fitting. Negative values of $\Delta$AIC and $\Delta$BIC suggest that the model under investigation performs better than the reference model. For positive values of $\Delta$AIC and $\Delta$BIC, we adopted the judgment criteria of the literature \citep{2020MNRAS.494.2576C,2022A&A...661A..71H}:
	\begin{itemize}\label{AICintervals}
		\item $\Delta\text{AIC(BIC)}\in [0,2]$ indicates weak evidence in favor of the reference model, leaving the question on which model is the most suitable open;
		\item $\Delta\text{AIC(BIC)}\in (2,6]$ indicates mild evidence against the given model with respect to the reference paradigm;
		and        \item $\Delta\text{AIC(BIC)}> 6$ indicates strong evidence against the given model, which should be rejected.
	\end{itemize}
	
	\section{Hemisphere comparison method} \label{method}
	The hemisphere comparison method proposed by \citet{2007A&A...474..717S} has been widely used in the investigation of the cosmic anisotropy, such as the anisotropy of cosmic expansion \citep{2018EPJC...78..755D,2022MNRAS.511.5661Z}, the acceleration scale of modified Newtonian dynamics \citep{2017ApJ...847...86Z,2018ChPhC..42k5103C}, and the temperature anisotropy of the CMB \citep{2004MNRAS.354..641H,2013ApJS..208...20B,2016JCAP...01..046G,2021PhRvD.104f3503F}. Firstly, we briefly introduce this method. Its goal is to identify the direction, which corresponds to the axis of maximal asymmetry from the dataset, by comparing the accelerating expansion rate ($q_{0}$). In the Pad$\rm \acute{e}$ approximations, it is convenient to employ $H_{0}$ to replace the accelerating expansion rate considering the relationship between the deceleration parameter $q_{0}$ and $H_{0}$. The most important step is to generate random directions $\hat{D}$ $(l$, $b)$ to divide the SNe dataset into two subsample (defined as "up" and "down"), where $l\in(0^{\circ}$, $360^{\circ})$ and $b\in(-90^{\circ}$, $90^{\circ})$ are the longitude and latitude in the galactic coordinate system, respectively. According to "up" and "down" subsamples, the corresponding best-fits of cosmological parameters are obtained employing the Markov chain Monte Carlo (MCMC) method. The anisotropy level (AL) made up of $q_{0,u}$ and $q_{0,d}$ can be used to describe the degree of deviation from isotropy. Its values can be derived from 
	\begin{equation}
		AL = \frac{\triangle q_{0}}{\bar{q}_{0}} = 2 \times \frac{q_{0,u} - q_{0,d}}{q_{0,u} + q_{0,d}},
		\label{AL}
	\end{equation}
	where $q_{0,u}$ and $q_{0,d}$ are the best-fits of the "up" subsample and "down" subsample, respectively. These
	two subsamples are separated from the full SNe sample by a random direction $\hat{D}$$(l,b)$. The $1\sigma$ uncertainty $\sigma_{AL}$ is
	\begin{equation}
		\sigma_{AL} = \frac{\sqrt{\sigma^{2}_{q^{\textnormal{max}}_{0,u}} + \sigma^{2}_{q^{\textnormal{max}}_{0,d}}}}{q^{\textnormal{max}}_{0,u} + q^{\textnormal{max}}_{0,d}},
		\label{sAL}
	\end{equation}
    where, $\sigma^{2}_{q^{\textnormal{max}}_{0,u}}$ and $\sigma^{2}_{q^{\textnormal{max}}_{0,d}}$ are the 1$\sigma$ errors corresponding to $q^{\textnormal{max}}_{0,u}$ and $q^{\textnormal{max}}_{0,d}$. During the calculation, we repeated 10000 random directions $\hat{D}$$(l,b)$. It is note that if we prefer to employed other parameters (for example $H_{0}$) to find the preferred directions of cosmic anisotropy, it is convenient to replace parameter $q_{0}$ of Eqs. (\ref{AL}) and (\ref{sAL}). 
	
	In order to examine whether the discrepancy degree of the cosmological parameters from the Pantheon+ sample is consistent with statistical isotropy, we plan to carry out statistical isotropic analyses. To achieve this, we spread the original data set evenly across the sky. After that, we were able to obtain the $\rm AL_{max}$ for the isotropic dataset. Meanwhile, an additional isotropic analysis was also considered. We preserved the spatial inhomogeneity of real sample and then randomly distributed the real dataset, which randomly redistributed the distance moduli and redshift combination to real-data positions (RP) only. Given the limitations of computing time, we repeated it 1000 times; this gave acceptable statistics. For convenience, we refer to these two approaches as isotropy analysis and isotropy RP analysis.
	
	\begin{table*}\footnotesize 
		\caption{Best fitting results at the 68 per cent confidence level using the Pantheon+ sample under different expansion methods and cosmological models.  \label{T1}}
		\centering
		\begin{spacing}{1.3}
			\begin{tabular}{lccccccccc} 
				\hline\hline
				Model & $H_{0}$  & $\Omega_{m}$ & $\Omega_{\Lambda}$ & $w_{0}$ & $w_{a}$ & $q_{0}$ & $j_{0}$ & $s_{0}$ & $l_{0}$ \\
				&  (km s$^{-1}$ Mpc$^{-1}$)  &  &  & &  &  &  & & \\
				\hline
				flat $\Lambda$CDM & 72.84$\pm$0.23  & 0.36$\pm$0.02 & -- & -- & -- & -- & --  & -- & -- \\
				non-flat $\Lambda$CDM & 72.53$_{-0.27}^{+0.28}$  & 0.26$_{-0.05}^{+0.06}$ & 0.48$\pm$0.08 & -- & -- & -- & -- & -- & -- \\
				$w$CDM & 72.51$\pm$0.27  & 0.21$_{-0.08}^{+0.10}$ & -- & $-$0.72$_{-0.10}^{+0.14}$ & -- & -- & -- & -- & -- \\
				CPL & 71.95$_{-0.42}^{+0.38}$  & 0.48$_{-0.04}^{+0.08}$ & -- & $-$0.52$_{-0.29}^{+0.20}$ & -7.6$_{-5.3}^{+5.4}$ & -- & --  & -- & -- \\
				$R_{h}$ = ct & 70.63$\pm$0.12  & -- & -- & -- & -- & -- & --  & -- & -- \\
				Pad$\rm \acute{e}$$_{(2,1)}$ & 72.53$\pm$0.28  & -- & -- & -- & -- & $-$0.35$_{-0.07}^{+0.08}$ & 0.43$_{-0.56}^{+0.38}$ & -- & -- \\
				Pad$\rm \acute{e}$$_{(2,1)}$ ($j_{0}$ = 1) & 72.74$\pm$0.22  & -- & -- & -- & -- & $-$0.43$\pm$0.03 & 1.0  &  -- & -- \\
				Pad$\rm \acute{e}$$_{(2,2)}$  & 72.48$^{+0.28}_{-0.29}$  & -- & -- & -- & -- & $-$0.33$^{+0.09}_{-0.09}$ & 0.13$^{+0.94}_{-0.68}$ & $-$0.56$^{+2.92}_{-0.98}$ & -- \\
				Pad$\rm \acute{e}$$_{(3,2)}$  & 72.33$^{+0.30}_{-0.30}$  & -- & -- & -- & -- & $-$0.22$^{+0.09}_{-0.09}$ & $-$1.58$^{+1.10}_{-0.78}$ & -- & -- \\
				\hline\hline
			\end{tabular}
		\end{spacing}
	\end{table*}
	
	
	\begin{table}\footnotesize
		\caption{Statistical results with respect to the reference flat $\Lambda$CDM model by utilizing of the AIC and BIC statistical criteria. \label{T2}}
		\begin{spacing}{1.3}
			\begin{tabular}{lccccc}
				\hline\hline
				Model & $\chi^{2}$ & AIC & $\Delta$AIC & BIC & $\Delta$BIC  \\ \hline
				flat $\Lambda$CDM            & 1753.0  & 1757.0 & 0.0 & 1767.9 & 0.0\\
				non-flat $\Lambda$CDM        & 1749.0  & 1755.0 & $-$2.0 & 1771.3 & 3.4\\
				$w$CDM                       & 1749.6  & 1755.6 & $-$1.4 & 1771.9 & 4.0\\
				CPL                          & 1748.3  & 1756.3 & $-$0.7 & 1778.1 & 10.2\\
				$R_{h}$ = ct                 & 1825.0  & 1827.0 & 70.0 & 1832.4 & 64.6\\
				Pad$\rm \acute{e}$$_{(2,1)}$               & 1748.7  & 1754.7 & $-$2.3 & 1771.0 & 3.1\\
				Pad$\rm \acute{e}$$_{(2,1)}$ ($j_{0}$ = 1) & 1750.3  & 1754.3 & $-$2.7 & 1765.2 & $-$2.7\\
				Pad$\rm \acute{e}$$_{(2,2)}$               & 1749.2  & 1757.2 &  0.2 & 1778.9 & 11.0\\
				\hline\hline
			\end{tabular}
		\end{spacing}
	\end{table}

	\section{Results and Discussions} \label{result}
	\subsection{Cosmographic constraints} \label{com}
\begin{figure}
		\centering
		\includegraphics[width=0.23\textwidth]{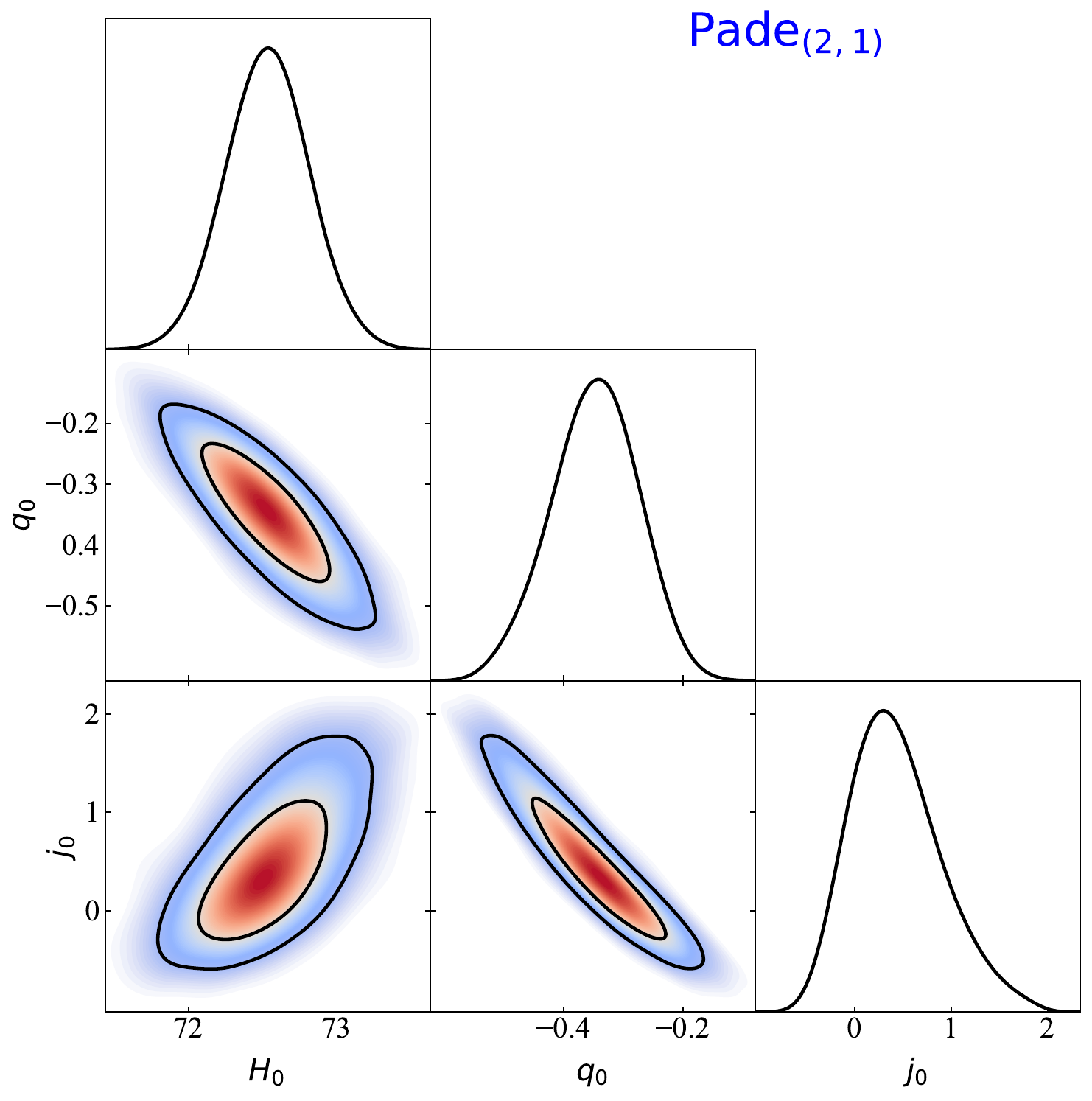}
		\includegraphics[width=0.23\textwidth]{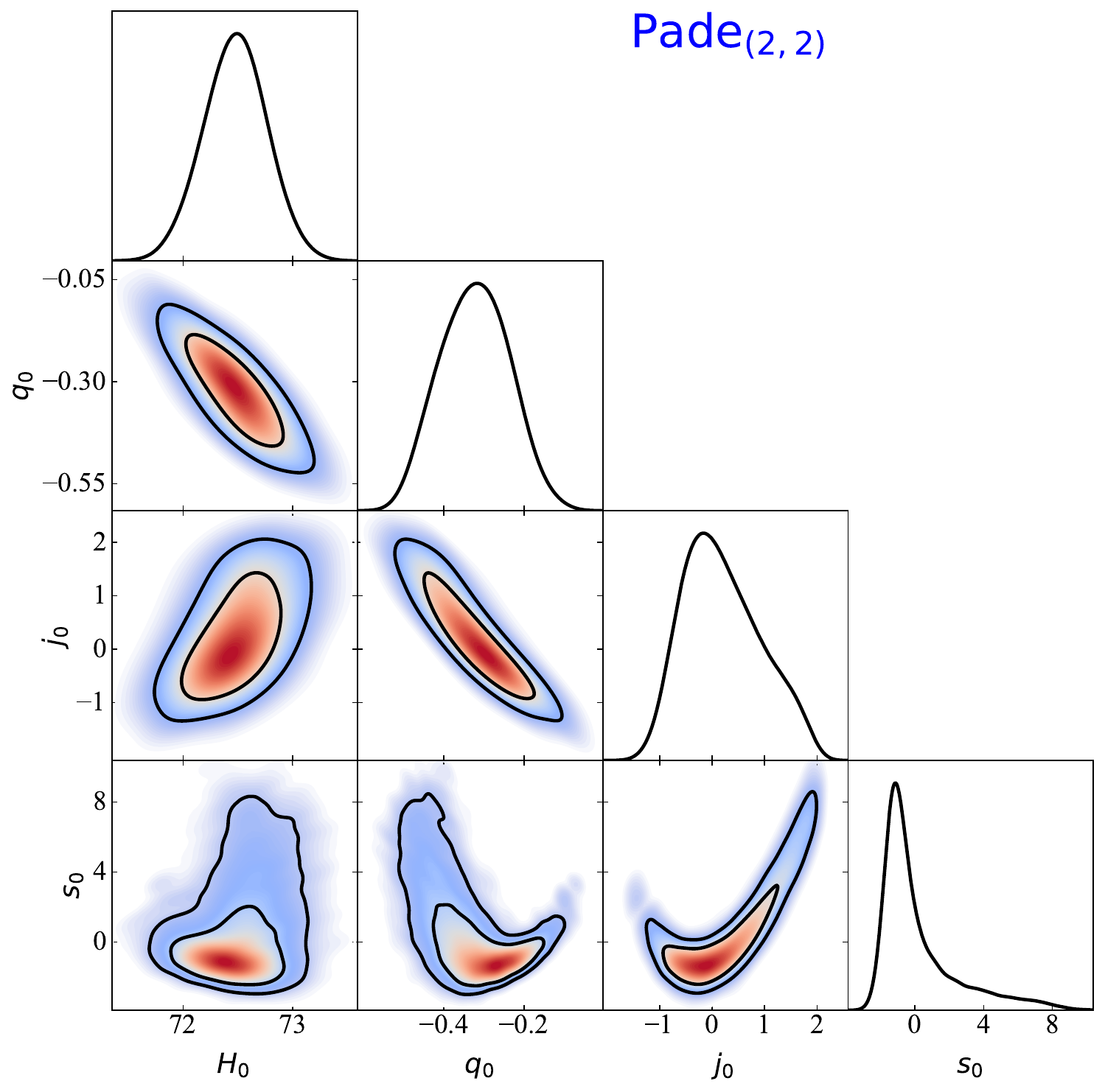} \\
		\includegraphics[width=0.23\textwidth]{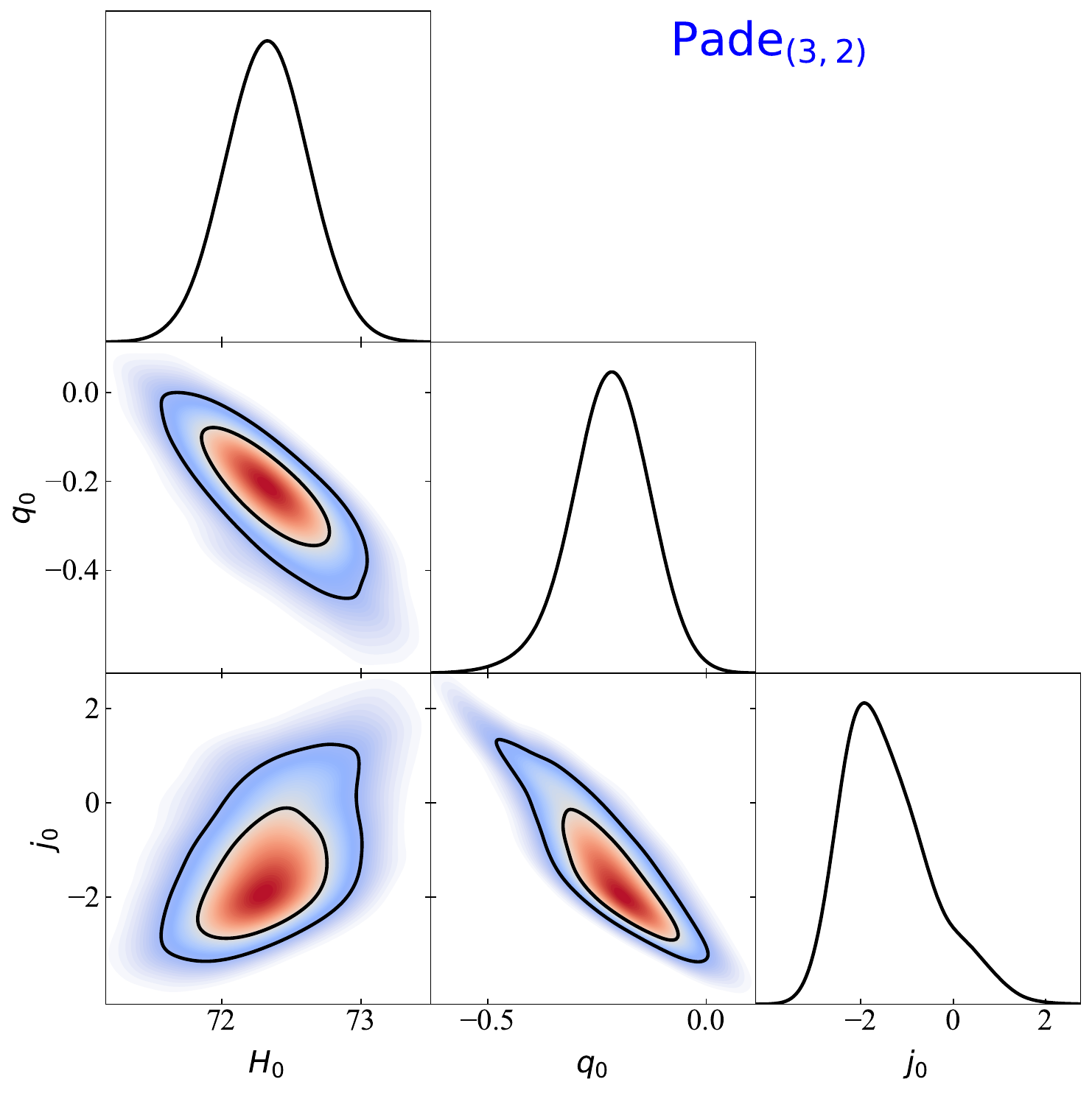}
		\includegraphics[width=0.23\textwidth]{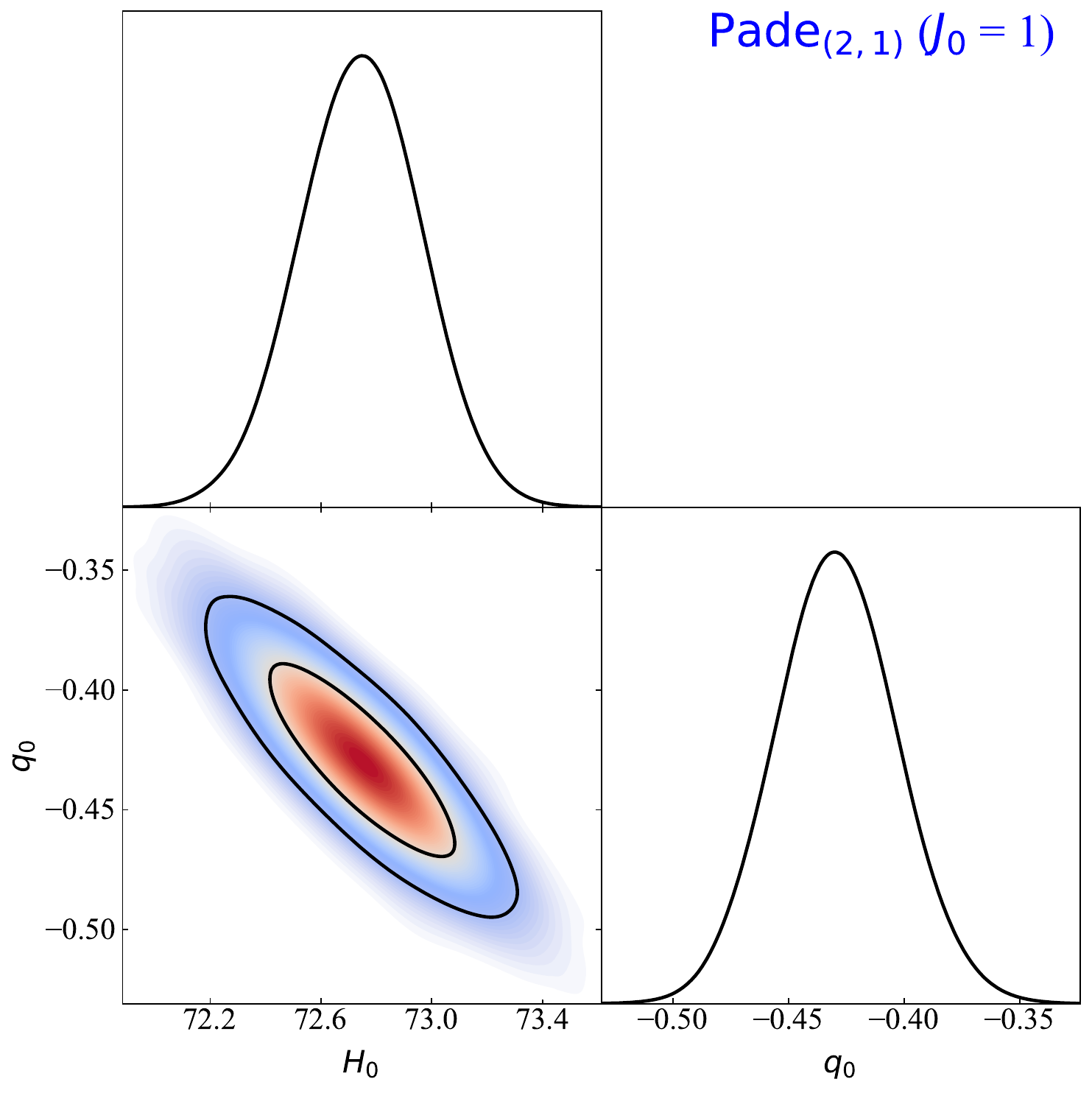}
		\caption{Confidence contours ($1\sigma$ and $2\sigma$) and marginalized likelihood distributions for the parameters space ($H_{0}$, $q_{0}$, $j_{0}$ and $s_{0}$) from the Pantheon+ sample utilizing the Pad$\rm \acute{e}$ approximations including Pad$\rm \acute{e}_{(2,1)}$, Pad$\rm \acute{e}_{(2,2)}$, Pad$\rm \acute{e}_{(3,2)}$ and Pad$\rm \acute{e}_{(2,1)}(j_{0} = 1)$ polynomials.}
		\label{F2}       
	\end{figure}

	We first gave the cosmographic constraints of the Pantheon+ sample by utilizing the Pad$\rm \acute{e}$ approximations, including the Pad$\rm \acute{e}_{(2,1)}$, Pad$\rm \acute{e}_{(2,2)}$, and Pad$\rm \acute{e}_{(3,2)}$ polynomials. The final contours are shown in Fig. \ref{F2}, and the detailed results are displayed in Table \ref{T1}. Overall, there are tight constraints for all parameters, except for the Pad$\rm \acute{e}_{(3,2)}$ polynomial. For the Pad$\rm \acute{e}_{(2,1)}$ polynomial, the results are $H_{0}$ = 72.53$\pm$0.28 km/s/Mpc, $q_{0}$ = $-$0.35$_{-0.07}^{+0.08}$, and $j_{0}$ = 0.43$_{-0.56}^{+0.38}$. For the Pad$\rm \acute{e}_{(2,2)}$ polynomial, the results are $H_{0}$ = 72.48$^{+0.28}_{-0.29}$ km/s/Mpc, $q_{0}$ = $-$0.33$^{+0.09}_{-0.09}$, $j_{0}$ = 0.13$^{+0.94}_{-0.68}$, $s_{0}$ = $-$0.56$^{+2.92}_{-0.98}$. For the Pad$\rm \acute{e}_{(3,2)}$ polynomial, the Pantheon+ sample could not give a tight constraint for all the parameters ($H_{0}$, $q_{0}$, $j_{0}$, $s_{0}$, $l_{0}$). To minimize risk, we only constrained the first two parameters ($q_{0}$ and $j_{0}$), and the other parameters were marginalized in a large range (0 $<$ $s_{0}$, $l_{0}$ $<$ 20) \citep{2017EPJC...77..434Z,2022A&A...661A..71H}. The final constraints are $H_{0}$ = 72.33$^{+0.30}_{-0.30}$ km/s/Mpc, $q_{0}$ = $-$0.22$^{+0.09}_{-0.09}$ and $j_{0}$ = $-$1.58$^{+1.10}_{-0.78}$. In the flat $\Lambda$CDM model, $q_{0}$ = 3$\Omega_{m}$/2 $-$ 1 and $j_{0}$ = 1 are expected. We can find that all $q_{0}$ and $j_{0}$ results are consistent with the $\Lambda$CDM model with $\Omega_{m}$ = 0.3 within 2$\sigma$ level. For $H_{0}$ constraints, they are in line with that from the SH0ES collaboration. The corresponding statistical information are shown in Table \ref{T2}. Since the Pad$\rm \acute{e}_{(3,2)}$ polynomial could not constrain all free parameters well, it should not be considered. The statistical results show that the Pantheon+ sample prefers the Pad$\rm \acute{e}_{(2,1)}$ polynomial in the Pad$\rm \acute{e}$ approximations. After that, we also gave the constraints of $H_{0}$ and $q_{0}$ employing the Pad$\rm \acute{e}_{(2,1)}$ polynomial with fixing $j_{0}$ = 1.0; that is $H_{0}$ = 72.74$\pm$0.22 km/s/Mpc and $q_{0}$ = $-$0.43$\pm$0.03. For convenience, it is referred to as Pad$\rm \acute{e}_{(2,1)}(j_{0} = 1)$. It is worth noting that although the $\chi^{2}$ value increases slightly, the values of AIC and BIC decrease significantly. This indicates that Pad$\rm \acute{e}_{(2,1)}(j_{0} = 1)$ polynomial is more suitable for Pantheon+ sample than the Pad$\rm \acute{e}_{(2,1)}$ polynomial.

	In addition, we also constrain the cosmological parameters combining the Pantheon+ sample with the usual cosmological models, such as the flat $\Lambda$CDM model, non-flat $\Lambda$CDM model, $w$CDM model, CPL model and $R_h$ = ct model. The corresponding confidence contours are shown in Figs. \ref{AF1} (flat and non-flat $\Lambda$CDM models), \ref{AF2} ($w$CDM and CPL models) and \ref{AF3} ($R_h$ = ct model). For the flat and non-flat $\Lambda$CDM model, the constraints are ($H_{0}$ = 72.84$\pm$0.23 km/s/Mpc, $\Omega_{m}$ = 0.36$\pm$0.02) and ($H_{0}$ = 72.53$_{-0.27}^{+0.28}$ km/s/Mpc, $\Omega_{m}$ = 0.26$_{-0.05}^{+0.06}$, $\Omega_{\Lambda}$ = 0.48$\pm$0.08), respectively. The value of $H_0$ is larger than that derived from $H(z)$ data \citep{Yu2018}. For the $w$CDM model, the results are $H_{0}$ = 72.51$\pm$0.27 km/s/Mpc, $\Omega_{m}$ = 0.21$_{-0.08}^{+0.10}$ and $w_{0}$ = $-$0.72$_{-0.10}^{+0.14}$. The results of CPL model are $H_{0}$ = 71.95$_{-0.42}^{+0.38}$ km/s/Mpc, $\Omega_{m}$ = 0.48$_{-0.04}^{+0.08}$, $w_{0}$ = $-$0.52$_{-0.29}^{+0.20}$ and $w_{a}$ = $-$7.6$_{-5.3}^{+5.4}$. The result $H_{0}$ = 70.63$\pm$0.12 km/s/Mpc is obtained using the $R_h$ = ct model. Finally, we made a comparison investigation between the Pad$\rm \acute{e}$ approximations and the usual cosmological models in terms of the AIC and BIC methods. The detailed information is shown in Table \ref{T2}. From the statistical information, we find that the Pad$\rm \acute{e}_{(2,1)}(j_{0} = 1)$ polynomials has better performance than the usual cosmological models. The smallest values for $\Delta$AIC and $\Delta$BIC were obtained by the Pad$\rm \acute{e}_{(2,1)}(j_{0} = 1)$ polynomials; that is $\Delta$AIC = $-$2.7 and $\Delta$BIC = $-$2.7. In addition, there exits negative value of $\Delta$AIC but with positive value of $\Delta$BIC, such as non-flat $\Lambda$CDM model, $w$CDM model, CPL model, and Pad$\rm \acute{e}_{(2,1)}(j_{0} = 1)$ polynomials. Overall Pad$\rm \acute{e}_{(2,1)}(j_{0} = 1)$ polynomial can describe the Pantheon+ sample well.

	\subsection{Cosmic anisotropy} 
\begin{figure}[h]
		\centering
		\includegraphics[width=0.45\textwidth]{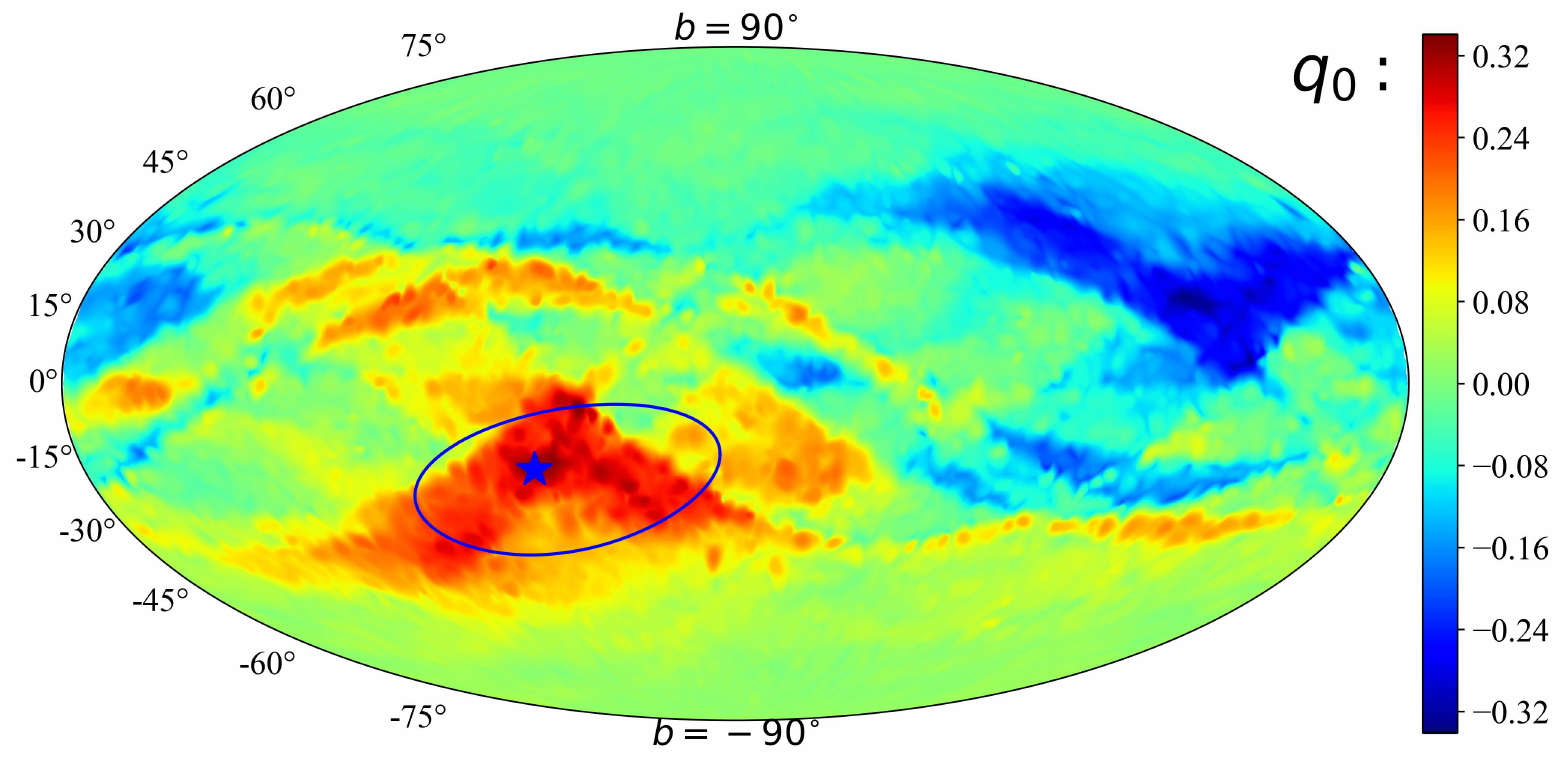}\\
		\includegraphics[width=0.45\textwidth]{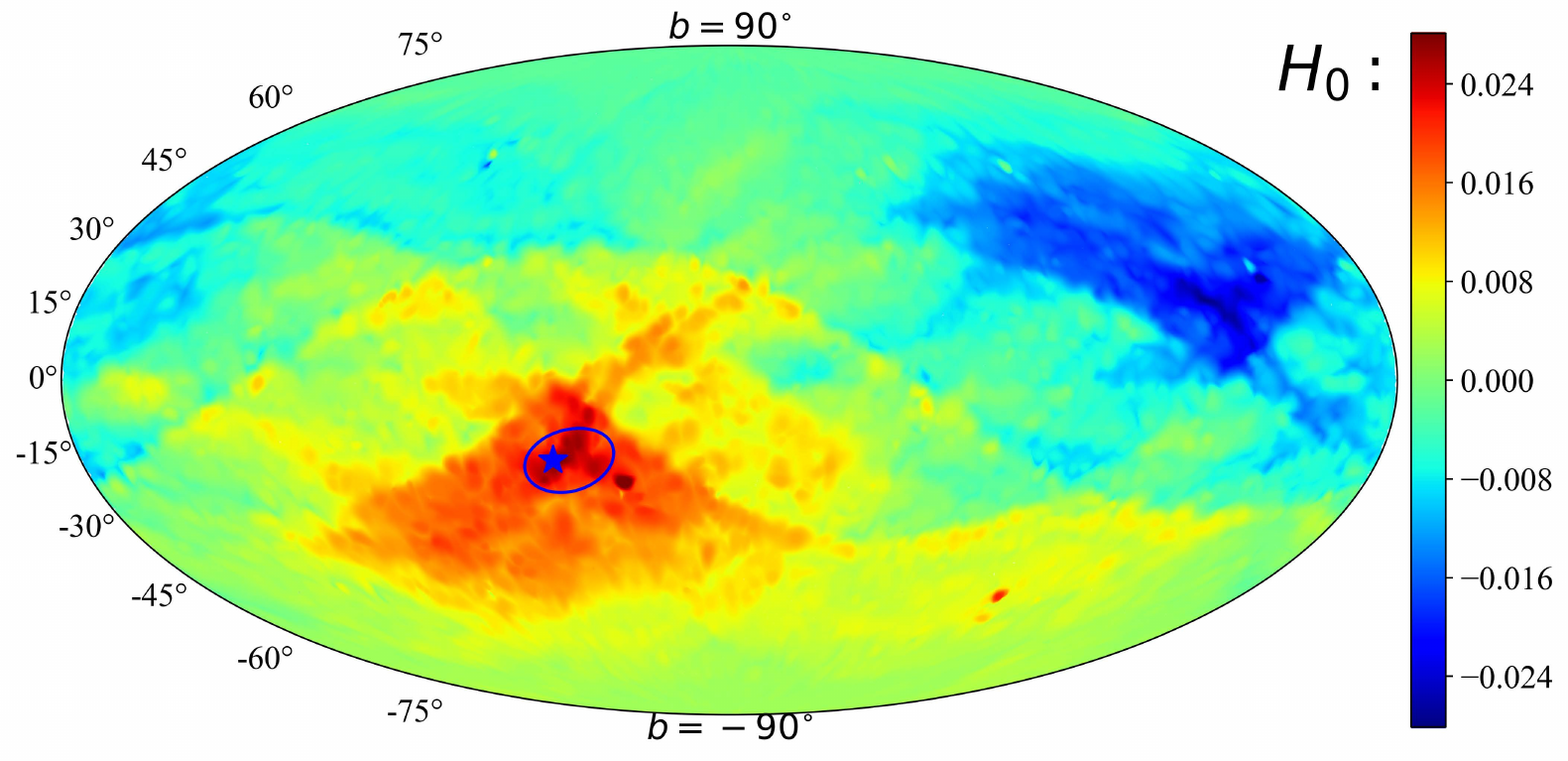}	
		\caption{Pseudo-color map of AL in the galactic coordinate system utilizing the HC method. The upper and lower panels show the results of $q_{0}$ and $H_{0}$, respectively. The blue star and line mark the direction of the largest AL and corresponding 1$\sigma$ range in the sky. The corresponding directions and 1$\sigma$ areas are parameterized as $\rm AL_{max, q_{0}}$ (304.6$^{\circ}$$_{-37.4}^{+51.4}$, $-$18.7$^{\circ}$$_{-20.3}^{+14.7}$) and $\rm AL_{max, H_{0}}$ (311.1$^{\circ}$$_{-8.4}^{+17.4}$, $-$17.53$^{\circ}$$_{-7.7}^{+7.8}$). }
		\label{F3}       
	\end{figure}

\begin{figure}
		\centering
		\includegraphics[width=0.23\textwidth]{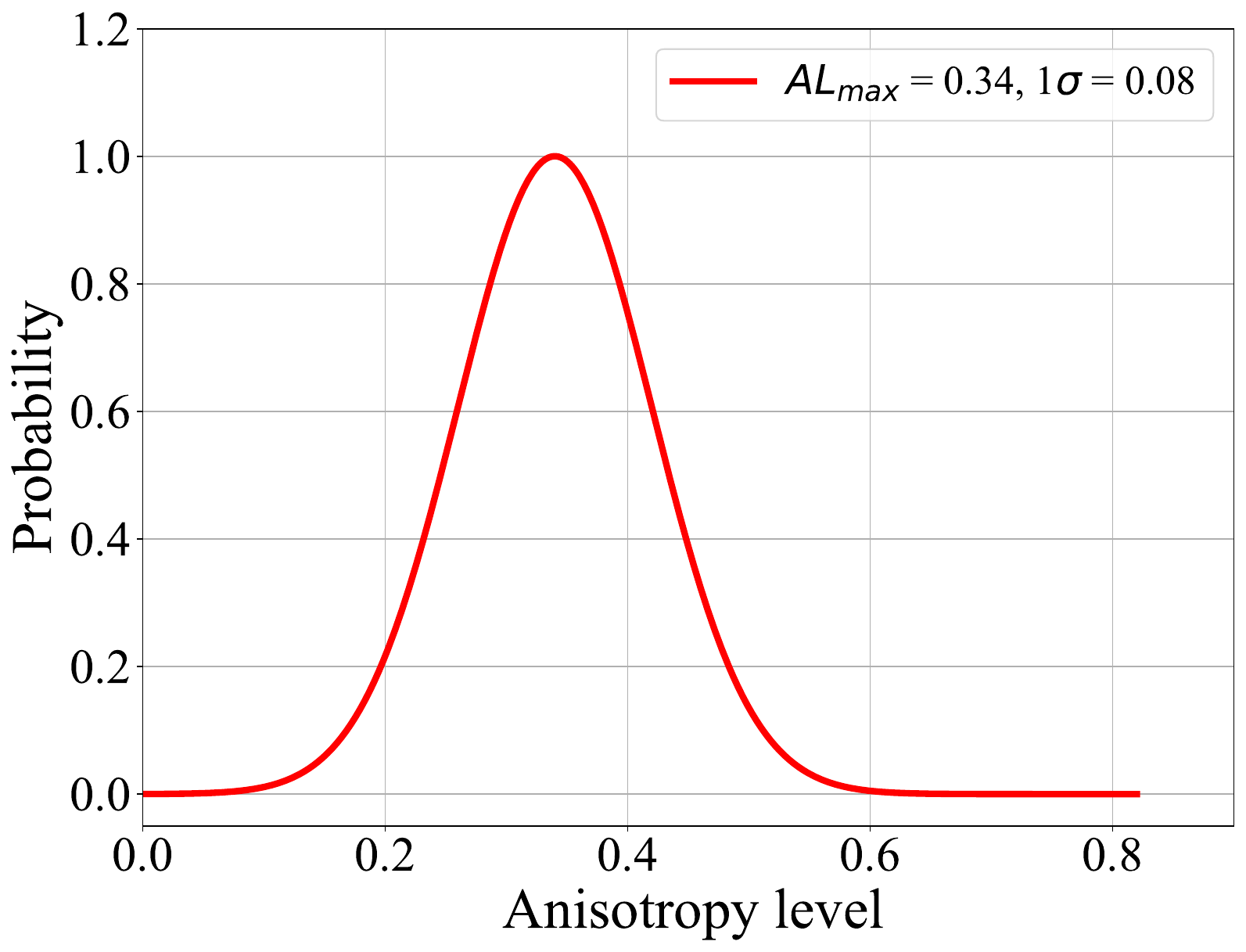}
		\includegraphics[width=0.237\textwidth]{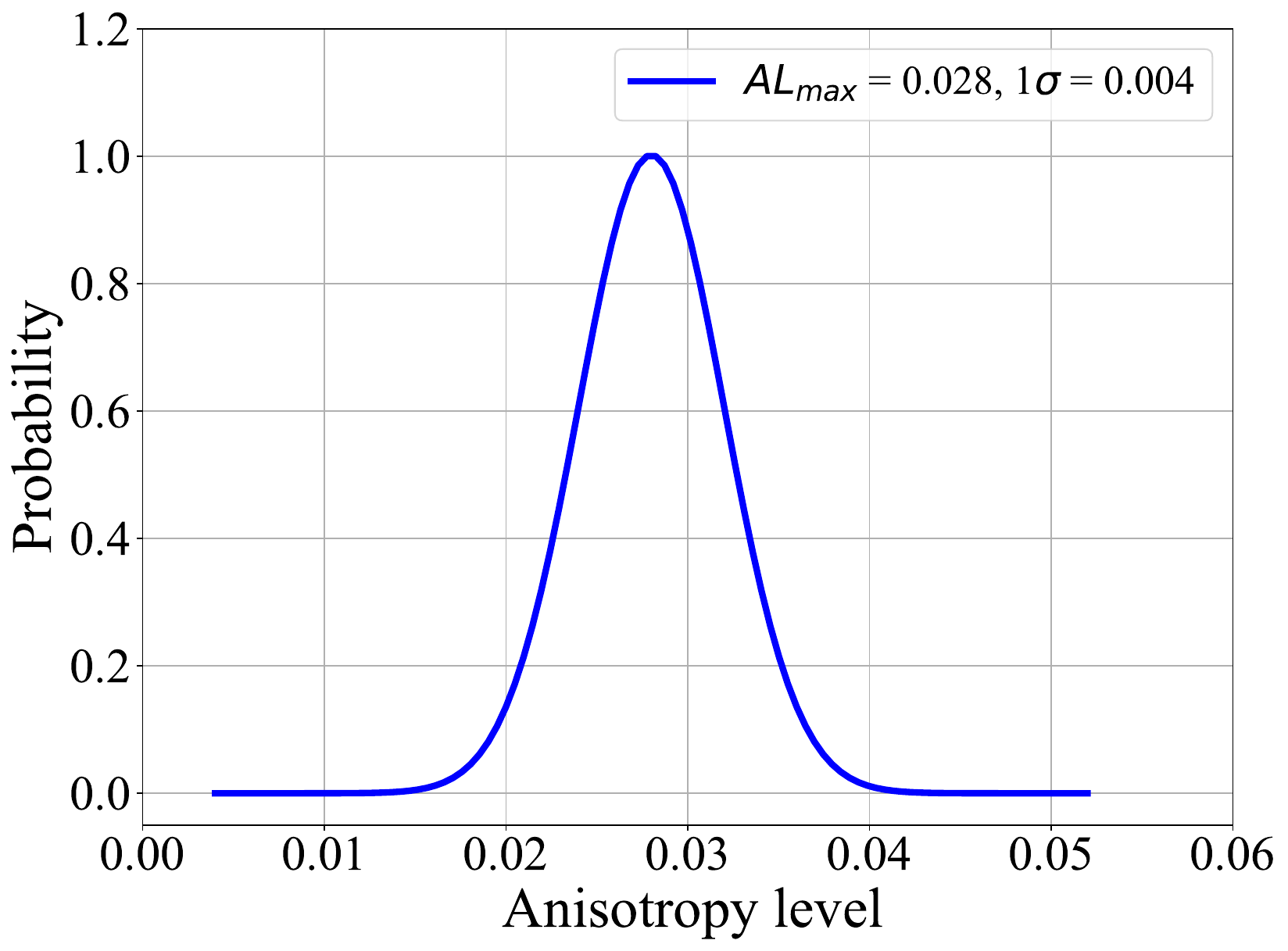}	
		\caption{Probability density distribution of the anisotropy level. The left and right panels are plotted by using $\rm AL_{max}(q_{0})$ = 0.34 $\pm$ 0.08 and $\rm AL_{max}(H_{0})$ = 0.028 $\pm$ 0.004), respectively. }
		\label{F31}       
	\end{figure}

	Based on the comparison analyses, we decided to combine the Pad$\rm \acute{e}_{(2,1)}(j_{0} = 1)$ polynomials with the HC method to find the preferred direction of cosmic anisotropy. Therefore, there exits two free parameters, $q_{0}$ and $H_{0}$. Previous work usually fixed the redundant parameters, leaving only one free parameter. Here, we chosen to fit $q_{0}$ and $H_{0}$ simultaneously for the different SNe subsamples which obtained by the random directions $\hat{D}$$(l,b)$. Utilizing Eq. (\ref{AL}), we mapped the corresponding AL$(l$, $b)$ distributions of $q_{0}$ and $H_{0}$ in the galactic coordinate system, as shown in Fig. \ref{F3}. From the AL distribution of $q_{0}$ (upper panel), we find there is a maximum value of AL, $\rm AL_{max}(q_{0})$ = 0.34, in direction (l, b) = (304.6$^{\circ}$$_{-37.4}^{+51.4}$, $-$18.7$^{\circ}$$_{-20.3}^{+14.7}$). The corresponding 1$\sigma$ error of AL is derived from Eq. (\ref{sAL}) and used to plot the 1$\sigma$ region; that is $\sigma_{AL_{max}}$ = 0.08. For the parameter $H_{0}$, we find $\rm AL_{max}(H_{0})$ = 0.028, $\sigma_{AL_{max}}$ = 0.004 and (l, b) = (311.1$^{\circ}$$_{-8.4}^{+17.4}$, $-$17.53$^{\circ}$$_{-7.7}^{+7.8}$), as shown in lower panel of Fig. \ref{F3}. It is easy to cover that these two preferred directions are consistent with each other. But $\rm AL_{max}$ and the corresponding $\sigma_{AL_{max}}$ values of $q_{0}$ are larger than that of $H_{0}$. In other word, the investigation of $H_{0}$ gives a tighter constraint for the anisotropic direction, but its anisotropic degree is weaker than that given by $q_{0}$ investigation. This means that these two parameters might have different sensitivities on the cosmic anisotropy. According to the values of $\rm AL_{max}$ and $\sigma_{AL_{max}}$, we plotted the probability density distributions of $\rm AL_{max}$, as shown in Fig. \ref{F31}. It suggests that the results of Fig. \ref{F3} significant departure from isotropy (AL = 0).
	
\begin{figure}
		\centering
		\includegraphics[width=0.24\textwidth]{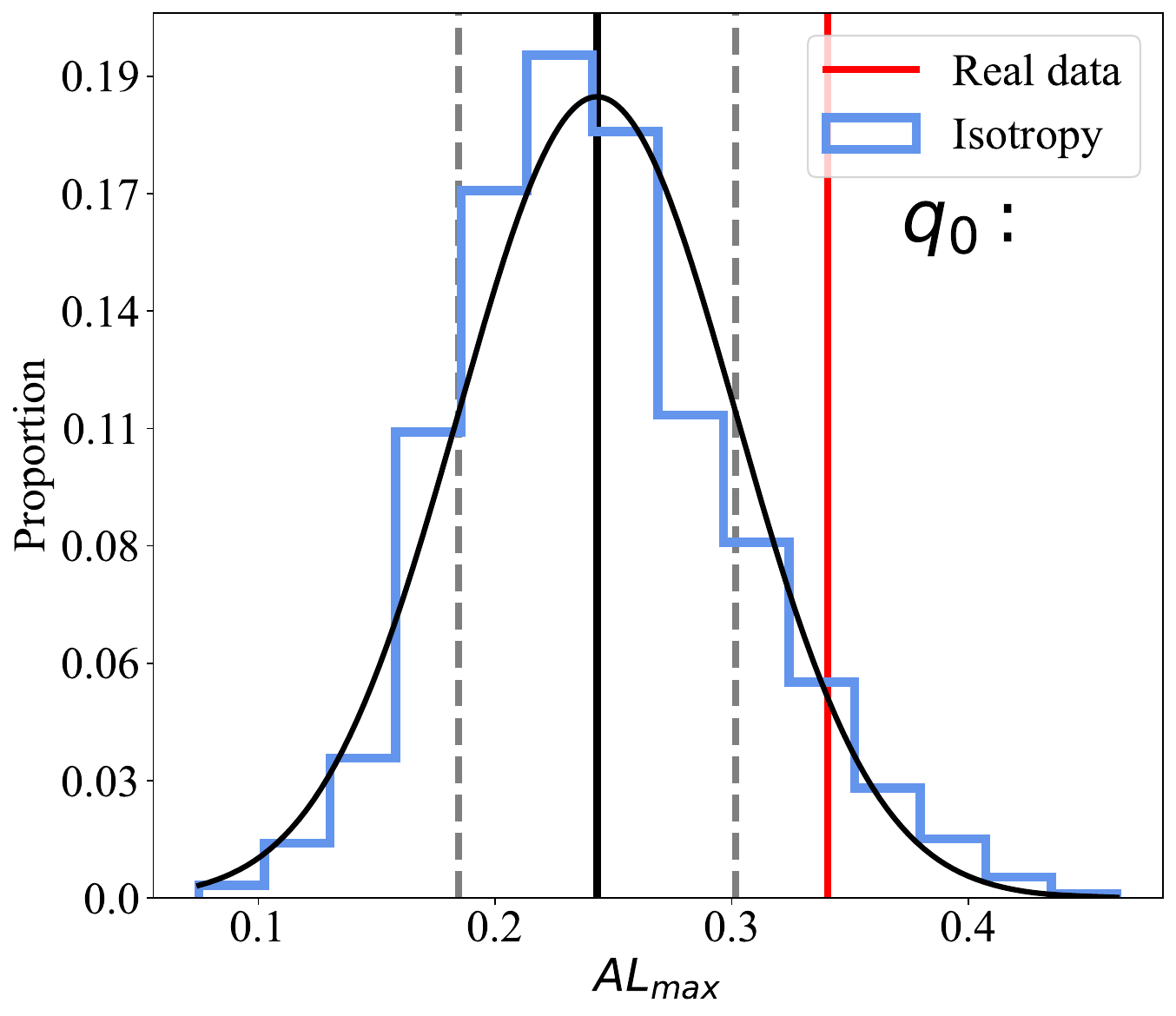} \
		\includegraphics[width=0.24\textwidth]{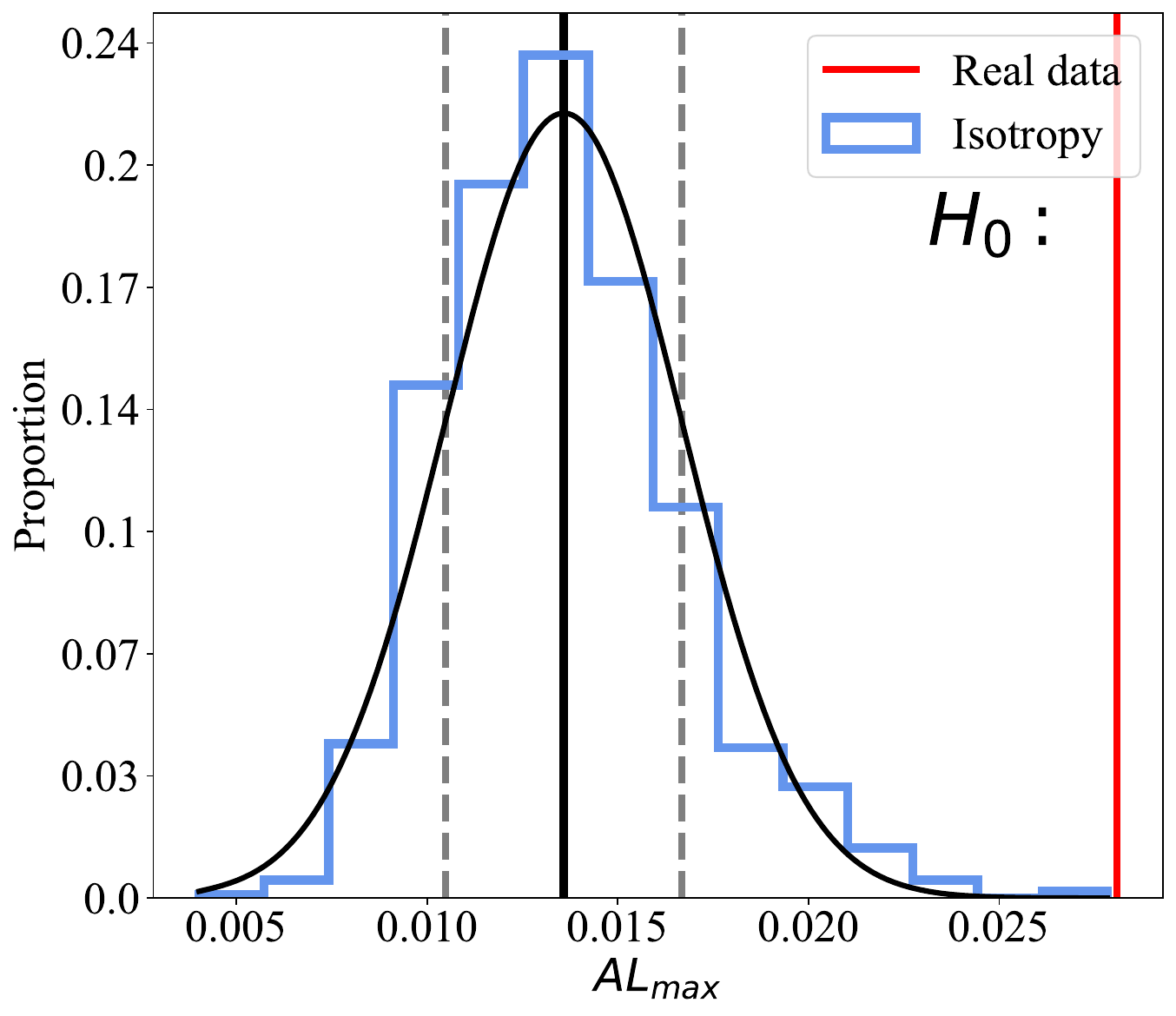} \\
		\includegraphics[width=0.24\textwidth]{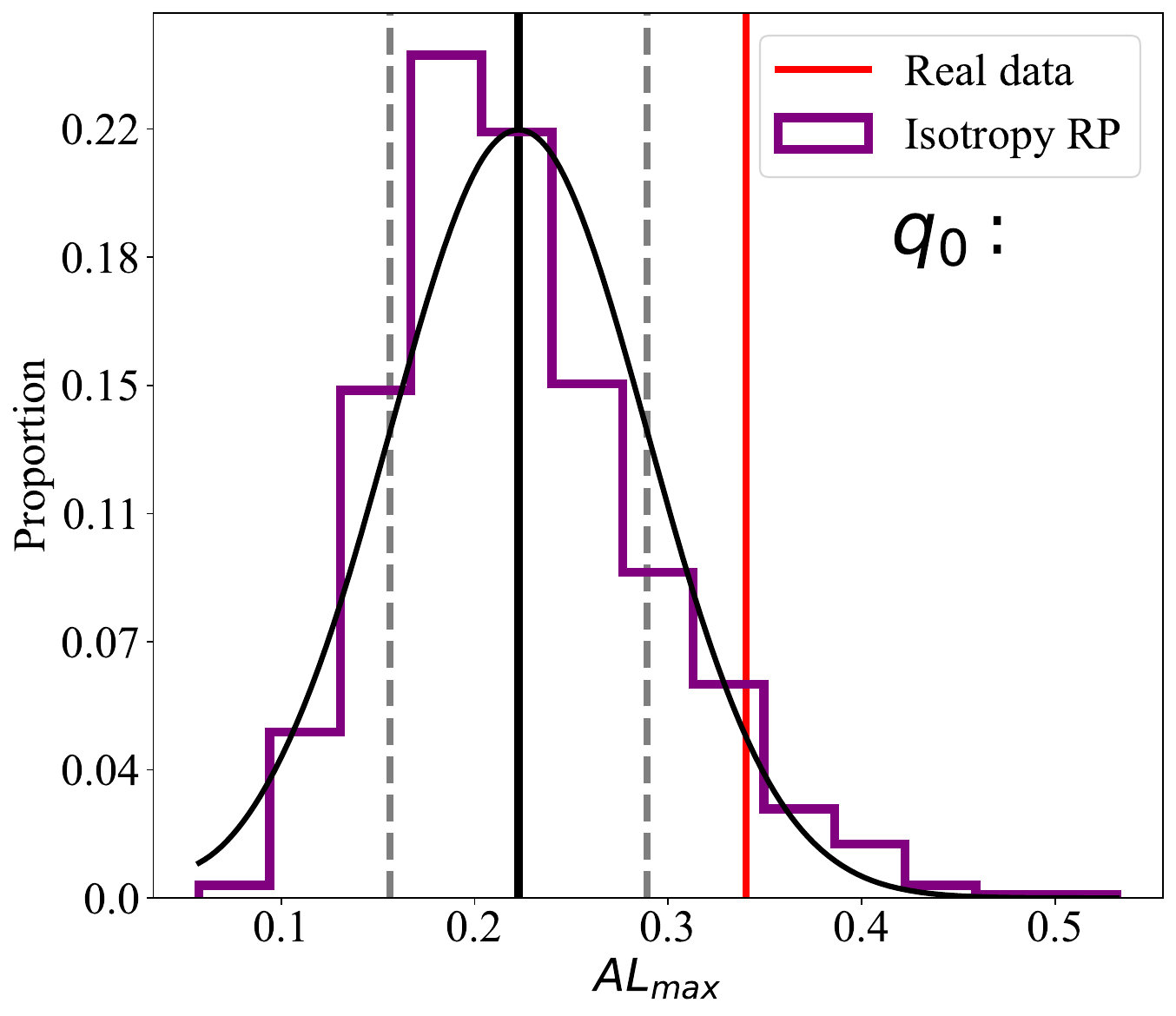} \ 
		\includegraphics[width=0.24\textwidth]{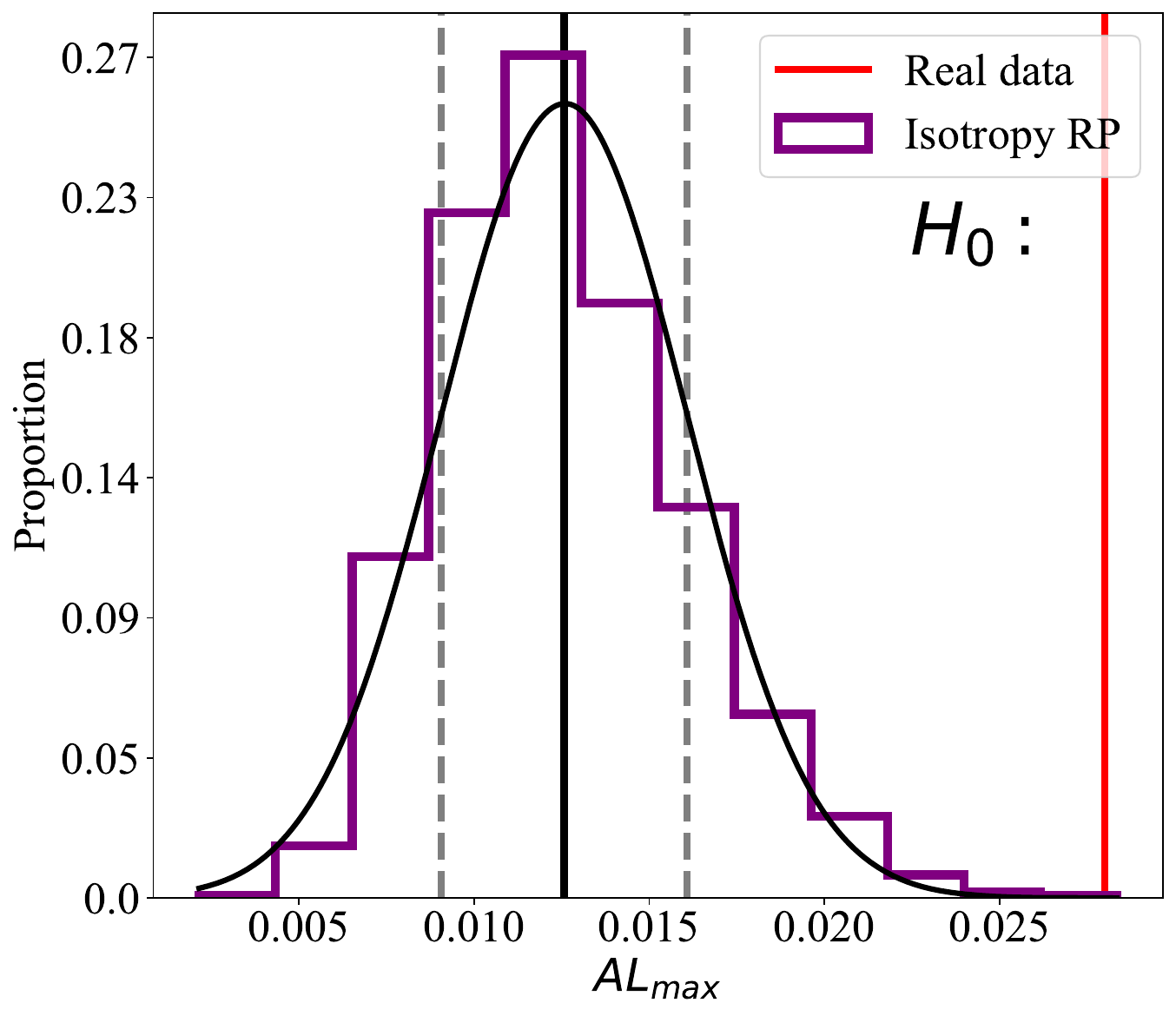}	
		\caption{Statistical results in 1000 simulated isotropic datasets. The Left and right panels show the results from $q_{0}$ and $H_{0}$ investigation, respectively. Colors blue and purple represent the statistical results of isotropic analyses (isotropy) and isotropic analyses that preserve the spatial inhomogeneity of real data (isotropy RP), respectively. The black curve is the best fitting result to the Gaussian function. The solid black and vertical dashed lines are commensurate with the mean and the standard deviation, respectively. The vertical red line shows the $\rm AL_{max}$ derived from the real data. For the isotropy analyses, the statistical significance of the real data are 1.72$\sigma$ ($q_{0}$) and 4.75$\sigma$ ($H_{0}$). For the isotropy RP analyses, the statistical significance are 1.78$\sigma$ ($q_{0}$) and 4.39$\sigma$ ($H_{0}$). }
		\label{F4}       
	\end{figure}

\begin{figure}
		\centering
		\includegraphics[width=0.5\textwidth]{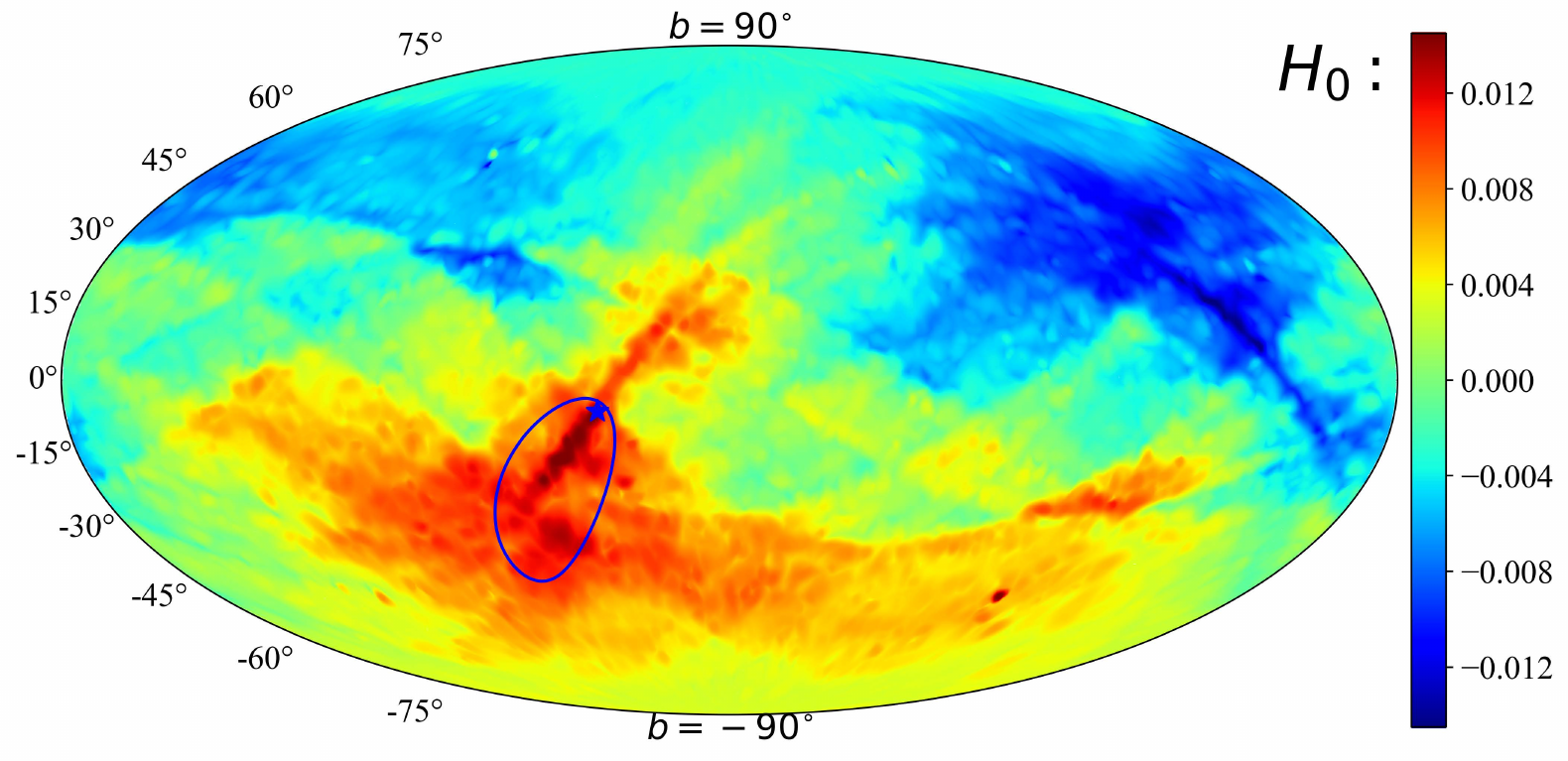}	
		\caption{Pseudo-color map of AL in the galactic coordinate system utilizing the HC method. The blue star and line mark the direction of the largest AL and corresponding 1$\sigma$ range in the sky. The direction and 1$\sigma$ area is parameterized as $\rm AL_{max, H_{0}}$ (324.3$^{\circ}$$_{-34.8}^{4.76}$, $-$6.7$^{\circ}$$_{-39.1}^{+3.0}$). Different from Fig. \ref{F3}, here only the parameter $H_{0}$ is free. Parameters $q_{0}$ and $j_{0}$ are fixed to $-$0.55 and 1.0, respectively. }
		\label{F5}       
	\end{figure}

\begin{figure}
		\centering
		\includegraphics[width=0.3\textwidth]{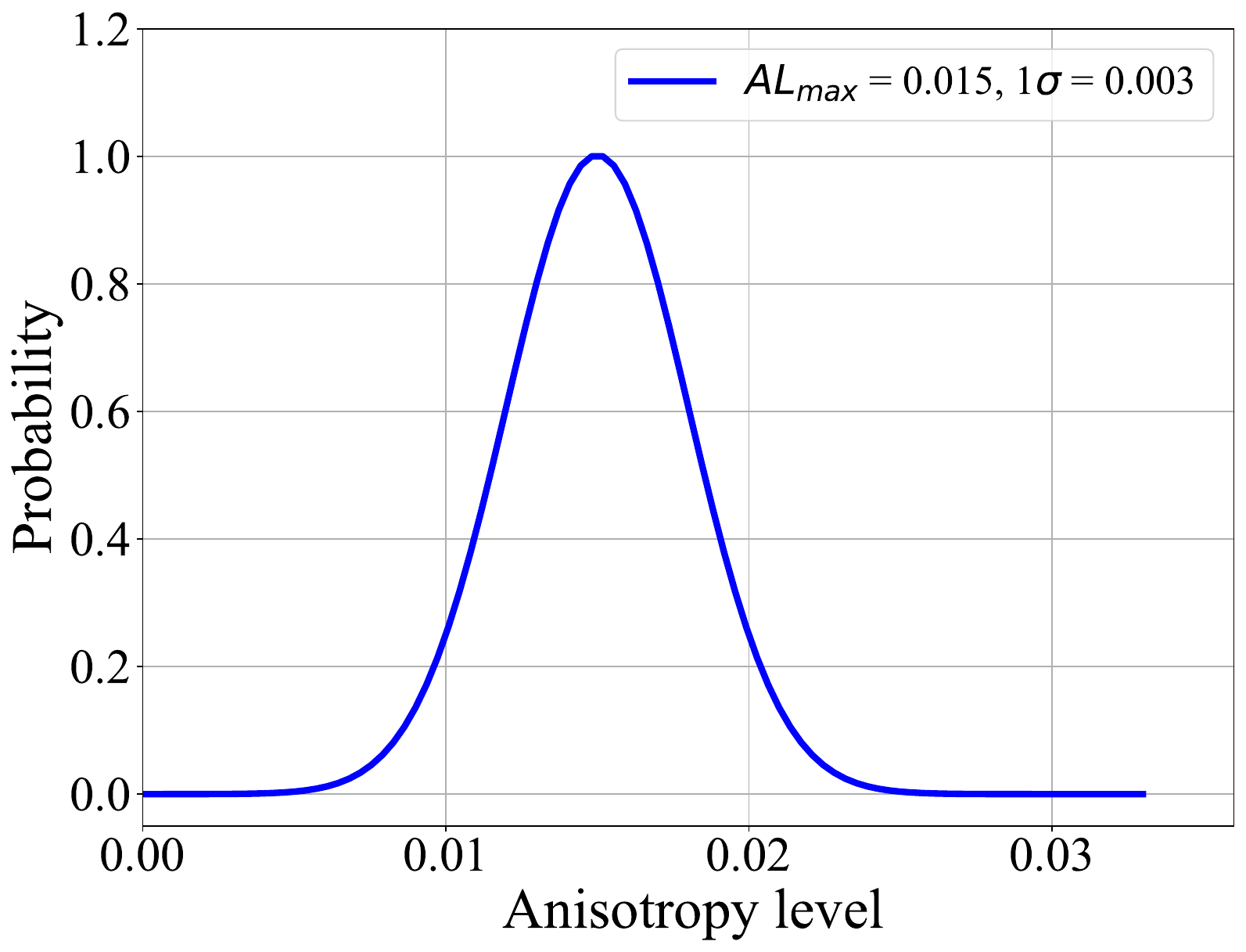}	
		\caption{Probability density distribution of the anisotropy level. The curve is constructed by $\rm AL_{max}(H_{0})$ = 0.015 $\pm$ 0.003.  }
		\label{F51}       
	\end{figure}

\begin{figure}
		\centering
		\includegraphics[width=0.24\textwidth]{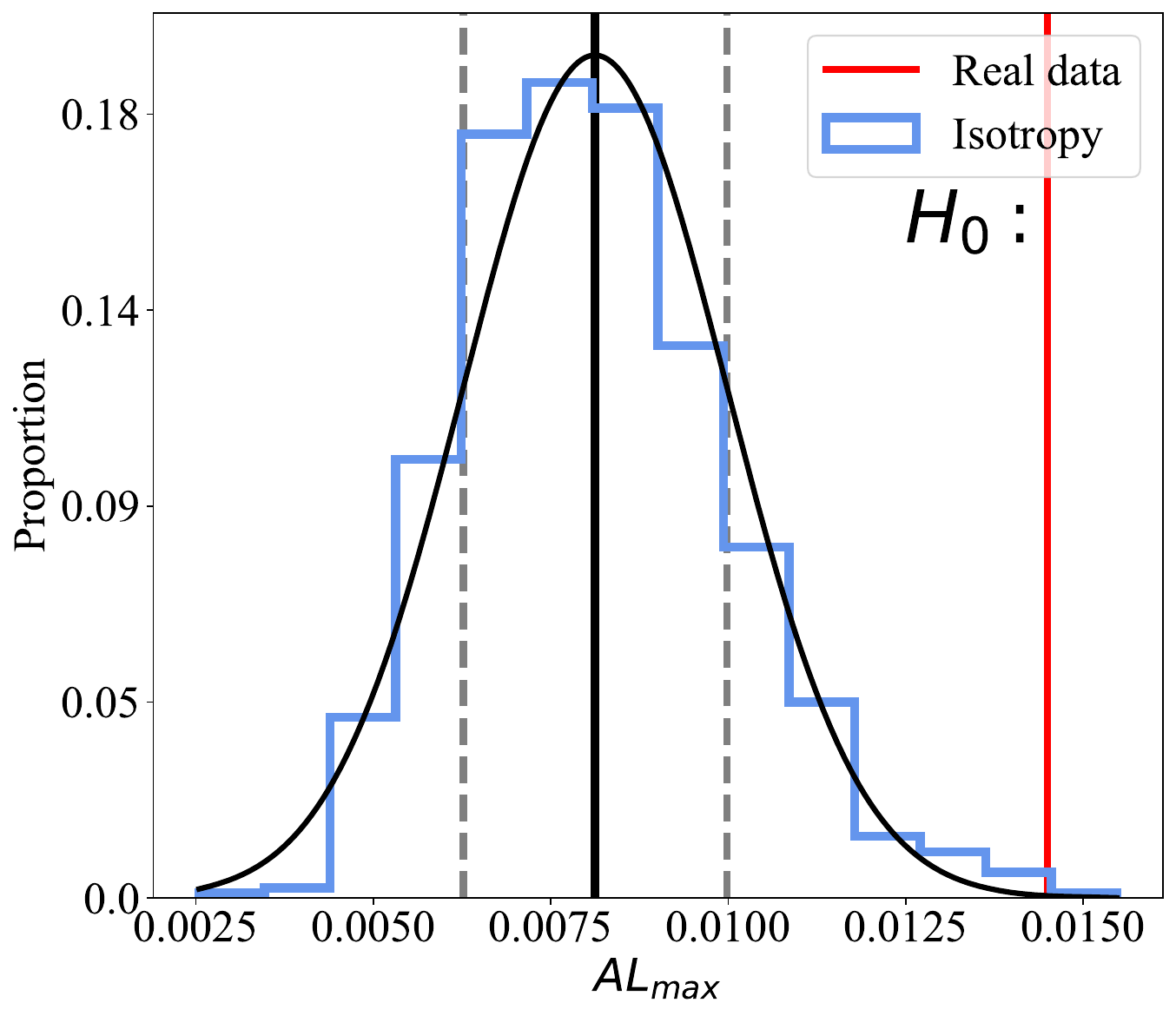} \
		\includegraphics[width=0.24\textwidth]{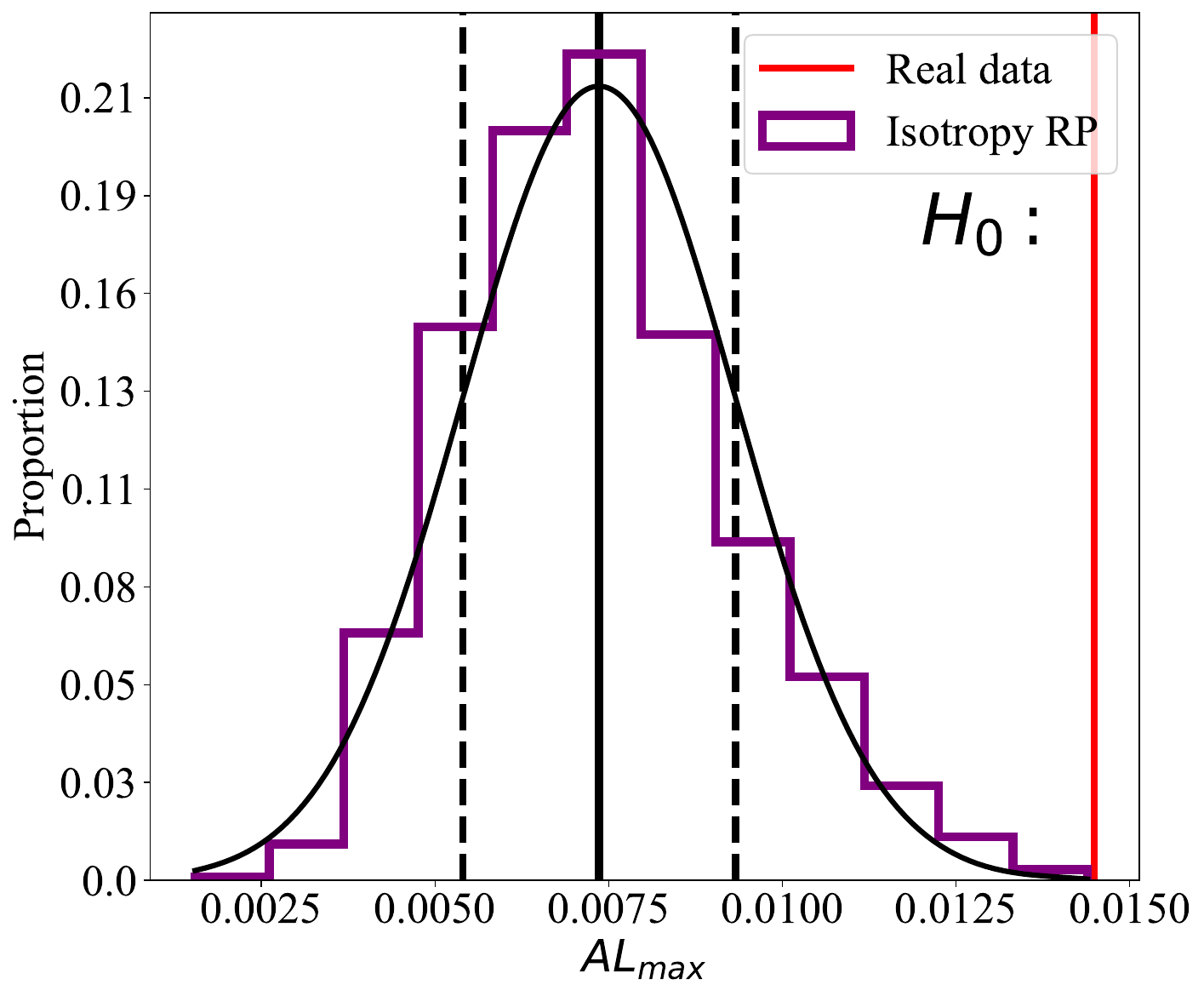}
		\caption{Statistical results in 1000 simulated isotropic datasets. Colors blue and purple represent the statistical results of isotropic analyses (isotropy) and isotropic analyses that preserve the spatial inhomogeneity of real data (isotropy RP); that is 3.42$\sigma$ and 3.62$\sigma$ respectively. The black curve is the best fitting result to the Gaussian function. The solid black and vertical dashed lines are commensurate with the mean and the standard deviation, respectively. The vertical red line shows the $\rm AL_{max}$ derived from the real data.}
		\label{F6}       
	\end{figure}

	The statistical isotropic analyses including the isotropy and isotropy RP can be well described by the Gaussian functions, as shown in Fig. \ref{F4}. For the isotropy analyses, the statistical significances of the real data are 1.72$\sigma$and 4.75$\sigma$ for $q_{0}$ and $H_{0}$, respectively. The statistical significances of the real data derived from the isotropy RP analyses are 1.78$\sigma$ for $q_{0}$ and 4.39$\sigma$ for $H_{0}$. From the statistical results, we can find that the statistical significances of $H_{0}$ are more obvious than that of $q_{0}$. This might be caused by the sensitivity differences of parameters on the cosmic anisotropy. The isotropy analyses give more obvious statistical significances than that given by the isotropy RF analyses, especially the $H_{0}$ investigations. This means that the spatial inhomogeneity can provide additional contribution on the cosmic anisotropy. In addition, we also considered to fix $q_{0}$ = $-$0.55 which corresponds to $\Omega_{m}$ = 0.30 to make reanalyses. The corresponding results are shown in Figs. \ref{F5}, \ref{F51} and \ref{F6}. From the AL distribution as shown in Fig. \ref{F5}, we obtained that $\rm AL_{max}(H_{0})$ = 0.015 $\pm$ 0.003 and (l, b) = (324.3$^{\circ}$$_{-34.8}^{+4.76}$, $-$6.7$^{\circ}$$_{-39.1}^{+3.0}$). Combining with the results of Figs. \ref{F3} and \ref{F5}, it can be found that the preferred direction of Fig. \ref{F5} deviates from that of Fig. \ref{F3}, but consistent with each other within 1$\sigma$ range. In addition, it is interesting that the direction of $\rm AL_{max}(H_{0})$ in Fig. \ref{F6} is close to the boundary of the 1 $\sigma$ range, rather than somewhere in the middle. This finding suggests that the current anisotropy of the Universe might be the result of the combined effect of multiple factors. If it is caused by a single factor, $\rm AL_{max}(H_{0})$ should be in the middle. The statistical significance has dropped slightly, but still greater than 3$\sigma$, that is 3.42$\sigma$ for the isotropy analyses and 3.62$\sigma$ for the isotropy RF analyses. All in all, the reanalyse results are in line with the previous results of Figs. \ref{F3} and \ref{F4}. From the overall results, the AL distributions of $q_{0}$ and $H_{0}$ give results that deviate significantly from isotropy. The corresponding statistical isotropic analyses of $H_{0}$ show higher statistical significance than that of $q_{0}$, that is near by 4.0$\sigma$. This provides a strong signal which hints a breach in the cosmological principle.

\begin{figure}
		\centering
		\includegraphics[width=0.5\textwidth]{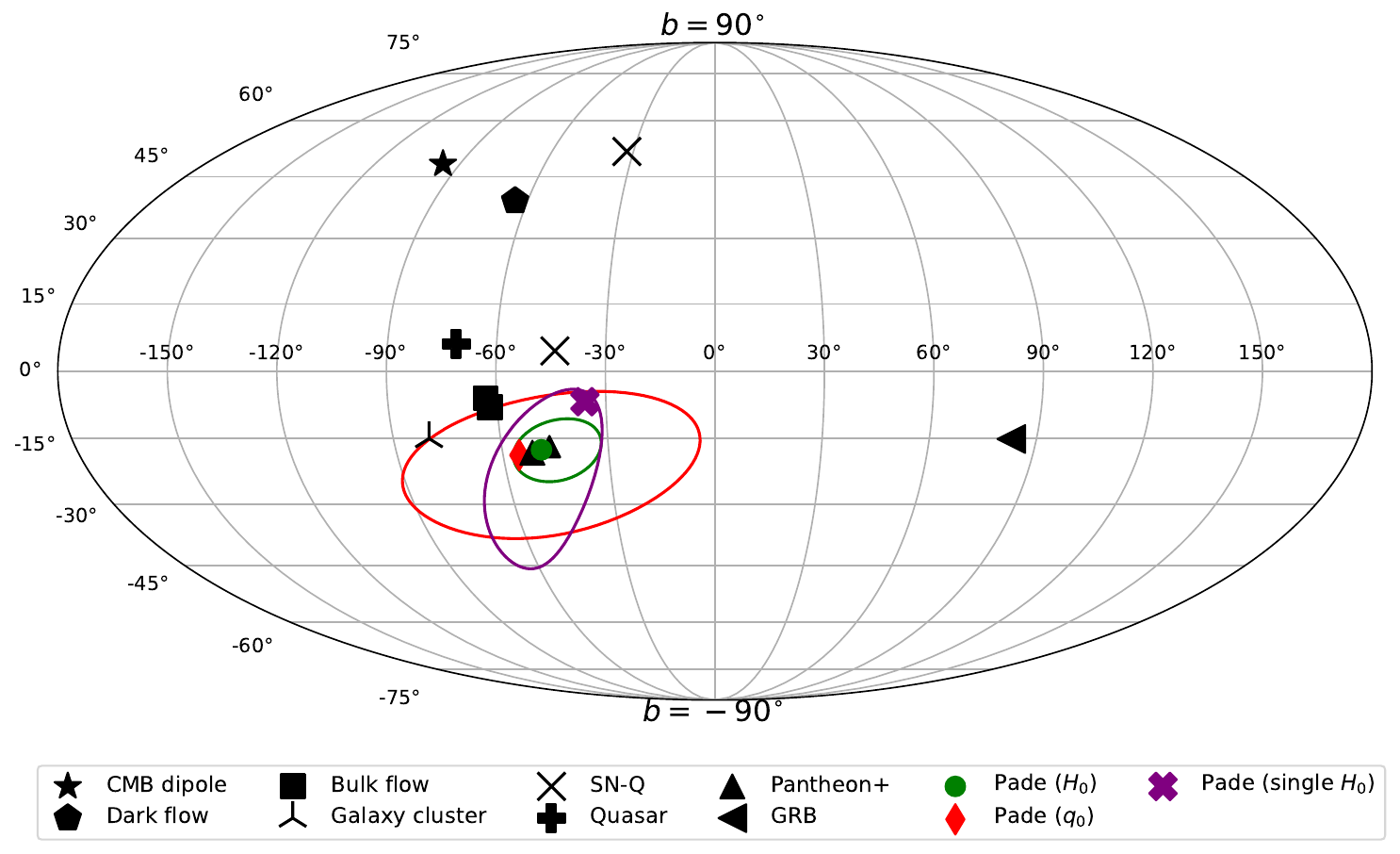}
		\caption{Distribution of our preferred directions (l, b) with 1$\sigma$ range. The colored marks represent the results we obtained. The black marks show the preferred directions after 2020 given by other observations including the CMB dipole \citep{2020AA...641A...1P}, dark flow \citep{2010ApJ...712L..81K}, bulk flow \citep{2023MNRAS.524.1885W}, galaxy cluster \citep{2020AA...636A..15M,2021AA...649A.151M}, SN-Q \citep{2020AA...643A..93H}, quasar \citep{2021EPJC...81..694Z}, quasar flux \citep{2024JCAP...06..019P}, Pantheon+ \citep{2024AA...681A..88H} and GRB \citep{2022MNRAS.511.5661Z}. }
		\label{F7}       
	\end{figure}

	All preferred directions we obtained are consistent with the results of previous researches that traced the anisotropy of $\Omega_{m}$ and $H_{0}$ \citep{2010JCAP...12..012A,2012JCAP...02..004C,2013AA...553A..56K,2015MNRAS.446.2952C,2020AA...643A..93H} and other dipole researches \citep{2014MNRAS.443.1680W,2014MNRAS.437.1840Y,2016MNRAS.456.1881L,2017MNRAS.468.1953P,2019MNRAS.486.5679Z,2023MNRAS.525..231D}, including SN Ia, quasar, GRB, and galaxy observations. However, they are different from those of \citet{2020PhRvD.102b3520K}, \citet{2022PhRvD.105j3510L}, \citet{2023PhRvD.108l3533M} and \citet{2024JCAP...06..019P}, which are consistent with the CMB dipole results \citep{2016A&A...594A...1P,2020AA...641A...1P}. Our results are also far from those obtained by \citet{2018PhRvD..97l3515D}, \citet{2018MNRAS.478.5153S} and \citet{2022MNRAS.511.5661Z}. In addition, comparing with other independent observations including the CMB dipole \citep{2016A&A...594A...1P,2020AA...641A...1P}, ultra-compact radio sources \citep{2012MNRAS.426..779J}, dark flow \citep{2010ApJ...712L..81K}, bulk flow \citep{2012MNRAS.420..447T,2017MNRAS.468.1420F,2023MNRAS.524.1885W}, and galaxy cluster \citep{2020AA...636A..15M,2021AA...649A.151M}, it is easy to find that the preferred directions we obtained are not consistent with the CMB dipole and ultra-compact radio sources, but they coincide with the galaxy cluster and the bulk flow. For ease of understanding, we aggregated the results after 2020 with the results we obtained and marked them on the galactic coordinate system, as shown in Fig. \ref{F7}. The more detailed information is shown in Table \ref{T3}. Anisotropy studies prior to 2020 can be found from Fig. 5 and Table 1 of \citet{2020AA...643A..93H}.

	\begin{table}\footnotesize
		\caption{Preferred directions given by observations. DF and HC are the abbreviations for the dipole fitting method and the hemisphere comparison method, respectively. \label{T3}}
		\centering
		\begin{spacing}{1.2}
			\begin{tabular}{ccc}
				\hline\hline
				Cosmological obs. & Direction (l, b)  & Ref.  \\
				\hline
				CMB dipole & (264.02$^{\circ}$$\pm$0.01, 48.25$^{\circ}$$\pm$0.01)  & (1)  \\
				Dark flow  &(296$^{\circ}$$\pm$28, 39$^{\circ}$$\pm$14)  & (2) \\
				Bulk flow  & (297$^{\circ}$$\pm$4, $-$4$^{\circ}$$\pm$3)  & (3) \\
				& (298$^{\circ}$$\pm$5, $-$7$^{\circ}$$\pm$4)  & (3) \\                 
				Galaxy cluster & (303$^{\circ}$, $-$27$^{\circ}$)  & (4) \\ 
				& (280$^{\circ}$$\pm$35, $-$15$^{\circ}$$\pm$20)  & (5) \\
				Quasar & (288.92$^{\circ}$$_{-28.80}^{+23.74}$, 6.10$^{\circ}$$_{-16.40}^{+16.55}$) & (6) \\
				Quasar flux & (201.50$^{\circ}$$\pm$27.87, $-$29.37$^{\circ}$$\pm$19.86) & (7) \\
				SN-Q (HC) & (316.08$^{\circ}$ $^{+27.41}_{-129.48}$, 4.53$^{\circ}$ $^{+26.29}_{-64.06}$) & (8) \\
				SN-Q (DF) & (327.55$^{\circ}$$\pm$32.45, 51.01$^{\circ}$$\pm$26.50) & (8) \\
				GRB & (82.97$^{\circ}$$_{-61.88}^{+52.73}$, $-$15.09$^{\circ}$$_{-13.54}^{+60.09}$) & (9) \\
				Pantheon & (286.93$^{\circ}$$\pm$18.52, 27.02$^{\circ}$$\pm$6.50) & (10) \\
				& (210.25$^{\circ}$$\pm$136.56, 72.85$^{\circ}$$\pm$60.63) & (10) \\
				Pantheon+ ($H_{0}$, 90$\degr$) & $({313.4^{\circ}}_{-18.2}^{+19.6}, {-16.8^{\circ}}_{-10.7}^{+11.1})$  & (11) \\
				Pantheon+ ($\Omega_{m}$, 90$\degr$) & (${308.4^{\circ}}_{-48.7}^{+47.6}, {-18.2^{\circ}}_{-28.8}^{+21.1}$)  & (11) \\
				\hline\hline
			\end{tabular}
			\begin{itemize}	
				\tiny
				\item[Ref.] (1) \citet{2020AA...641A...1P}, (2) \citet{2010ApJ...712L..81K}, (3) \citet{2023MNRAS.524.1885W}, (4) \citet{2020AA...636A..15M}, (5) \citet{2021AA...649A.151M}, (6) \citet{2021EPJC...81..694Z}, (7) \citet{2024JCAP...06..019P}, (8) \citet{2020AA...643A..93H}, (9) \citet{2022MNRAS.511.5661Z}, (10) \citet{2020PhRvD.102b3520K}, (11) \citet{2024AA...681A..88H}.
			\end{itemize}
		\end{spacing}
	\end{table}

	\section{Summary} \label{conl}
	In this paper, we combined the Pad$\rm \acute{e}$ approximation with the latest SNe Ia sample (Pantheon+) for cosmological researches. First, we gave the cosmographic constraints of third-, fourth- and fifth-order utilizing the Pad$\rm \acute{e}$ approximations, then found that the third order polynomial (Pad$\rm \acute{e}$ (2, 1)) can describe the Pantheon+ sample better. The corresponding results are $H_{0}$ = 72.53$\pm$0.28 km/s/Mpc, $q_{0}$ = $-$0.35$_{-0.07}^{+0.08}$, and $j_{0}$ = 0.43$_{-0.56}^{+0.38}$. The Pad$\rm \acute{e}_{(2,1)}$ polynomial with fixing $j_{0}$ = 1.0 can give tighter constraints than that of the regular Pad$\rm \acute{e}_{(2,1)}$ polynomial. The corresponding statistical results also show that the Pantheon+ sample prefers the Pad$\rm \acute{e}_{(2,1)}$$(j_{0} = 1.0)$ polynomial. The comparison investigation between the Pad$\rm \acute{e}$ approximations and the usual cosmological models shows that the Pad$\rm \acute{e}_{(2,1)}(j_{0} = 1)$ polynomials has better performance than the usual cosmological models including flat $\Lambda$CDM model, non-flat $\Lambda$CDM model, $w$CDM model, CPL model, and $R_h$ = ct model.
	
	Based on the Pad$\rm \acute{e}_{(2,1)}(j_{0} = 1)$ polynomial, we tested the cosmological principle utilizing the Pantheon+ sample and the HC method. Different from the previous works, we choose to fit parameters $H_{0}$ and $q_{0}$ simultaneously. We gave the preferred directions of cosmic anisotropy; that is (l, b) = (304.6$^{\circ}$$_{-37.4}^{+51.4}$, $-$18.7$^{\circ}$$_{-20.3}^{+14.7}$) for $q_{0}$ and  (311.1$^{\circ}$$_{-8.4}^{+17.4}$, $-$17.53$^{\circ}$$_{-7.7}^{+7.8}$) for $H_{0}$. The statistical isotropy analyses give the statistical significance of the real data are 1.72$\sigma$ and 4.75$\sigma$ for $q_{0}$ and $H_{0}$, respectively. The ones of isotropy RP analyses are 1.78$\sigma$ for $q_{0}$, and 4.39$\sigma$ for $H_{0}$. The reanalyses fixing $q_{0}$ = $-$0.55 give similar results. The preferred direction towards (l, b) = (324.3$^{\circ}$$_{-34.8}^{+4.76}$, $-$6.7$^{\circ}$$_{-39.1}^{+3.0}$), and the statistical significance are 3.42$\sigma$ and 3.62$\sigma$ for the isotropy analyses and isotropy RP analyses, respectively. Our obtained preferred directions are consistent with each other within 1$\sigma$ range, and they are in line with the results obtained from the different kinds of observations \citep{2010JCAP...12..012A,2012JCAP...02..004C,2012MNRAS.420..447T,2013AA...553A..56K,2014MNRAS.443.1680W,2014MNRAS.437.1840Y,2015MNRAS.446.2952C,2016MNRAS.456.1881L,2017MNRAS.468.1420F,2017MNRAS.468.1953P,2019MNRAS.486.5679Z,2020AA...643A..93H,2020AA...636A..15M,2021AA...649A.151M,2023MNRAS.525..231D,2023MNRAS.524.1885W}. Of course, there also exist some inconsistent results \citep{2012MNRAS.426..779J,2016A&A...594A...1P,2018PhRvD..97l3515D,2018MNRAS.478.5153S,2020PhRvD.102b3520K,2020AA...641A...1P,2022JHEAp..34...49A,2022PhRvD.105j3510L,2023PhRvD.108l3533M,2024JCAP...06..019P}. In short, the Pad$\rm \acute{e}$ approximation can be well combined with the HC method and applied to the study of cosmic anisotropy. The results show that relatively obvious anisotropic signals were found from the Pantheon+ sample. In order to better understand this signal, further analysis is still needed in the future.

	\section*{Acknowledgements}
	This work was supported by the National Natural Science Foundation of China (grant No. 12273009), the China Manned Spaced Project (CMS-CSST-2021-A12), Jiangsu Funding Program for Excellent Postdoctoral Talent (20220ZB59), Project funded by China Postdoctoral Science Foundation (2022M721561), Yunnan Youth Basic Research Projects (202001AU070013) and National Natural Science Foundation of China (grant No. 12373026).

	\bibliographystyle{aa} 

	\clearpage
	\appendix
	\section{1$\sigma$ and 2$\sigma$ contours in the 2-D parameter space utilizing different cosmological models}
	
	
	\begin{figure}[h]
		\centering
		\includegraphics[width=0.24\textwidth]{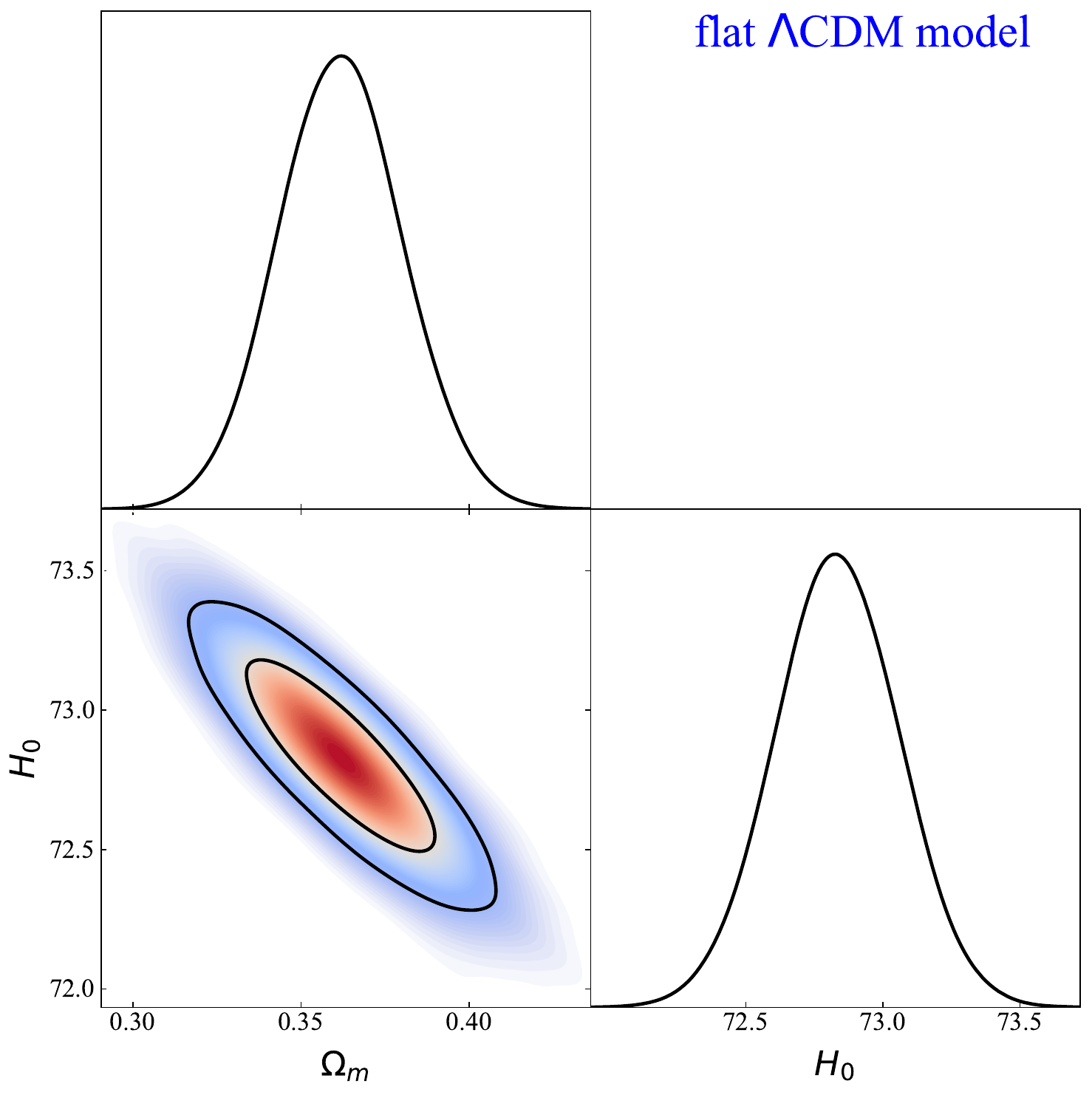} 
		\includegraphics[width=0.24\textwidth]{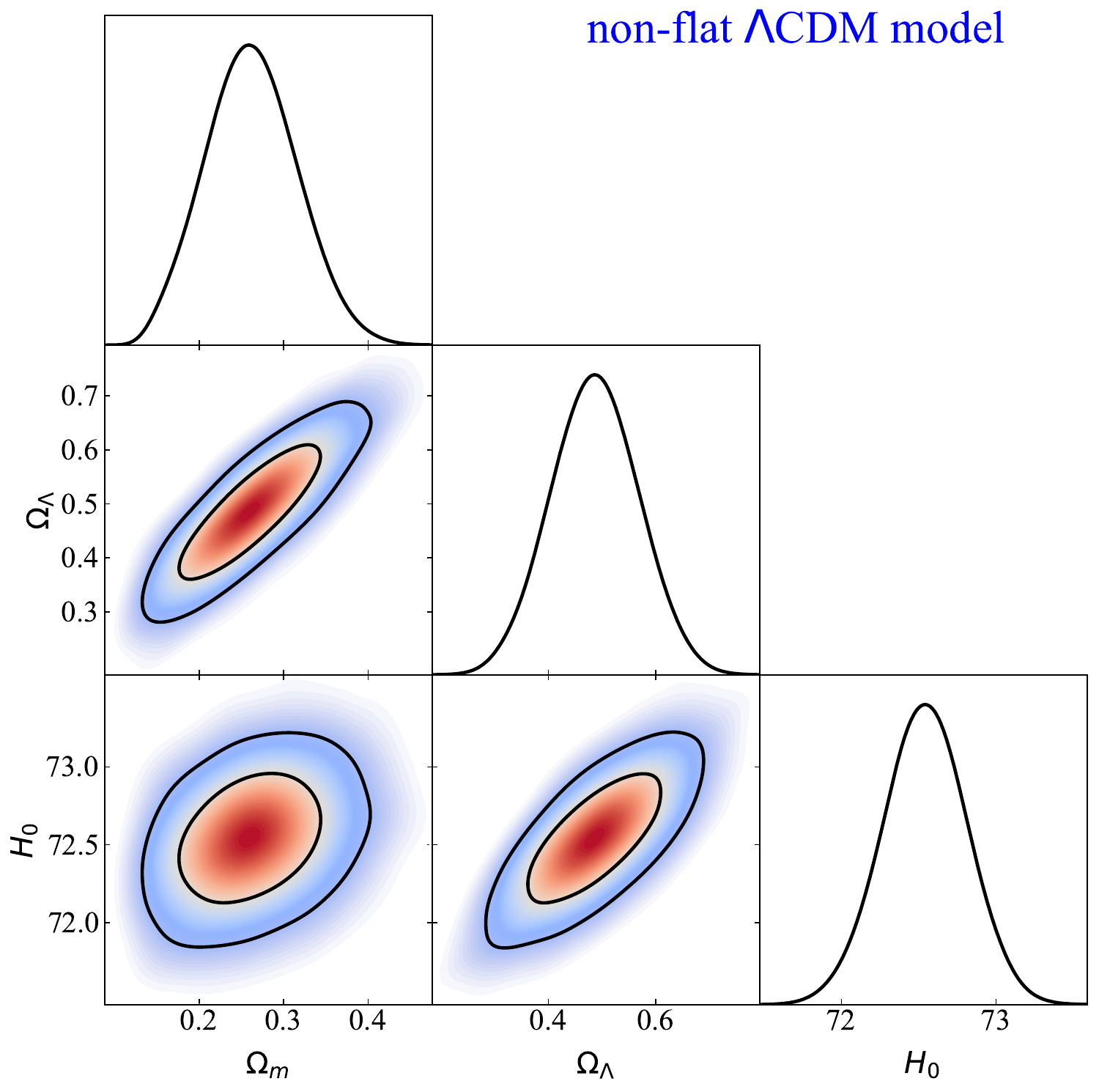}
		\caption{Confidence contours ($1\sigma$ and $2\sigma$) and marginalized likelihood distributions for the parameters space ($\Omega_{m}$, $\Omega_{\Lambda}$ and $H_{0}$) employing the Pantheon+ sample in the $\Lambda$CDM models. Left and right panels show the results from the flat and non-flat $\Lambda$CDM models, respectively.}
		\label{AF1}       
	\end{figure}

	
	\begin{figure}[h]
		\centering
		\includegraphics[width=0.24\textwidth]{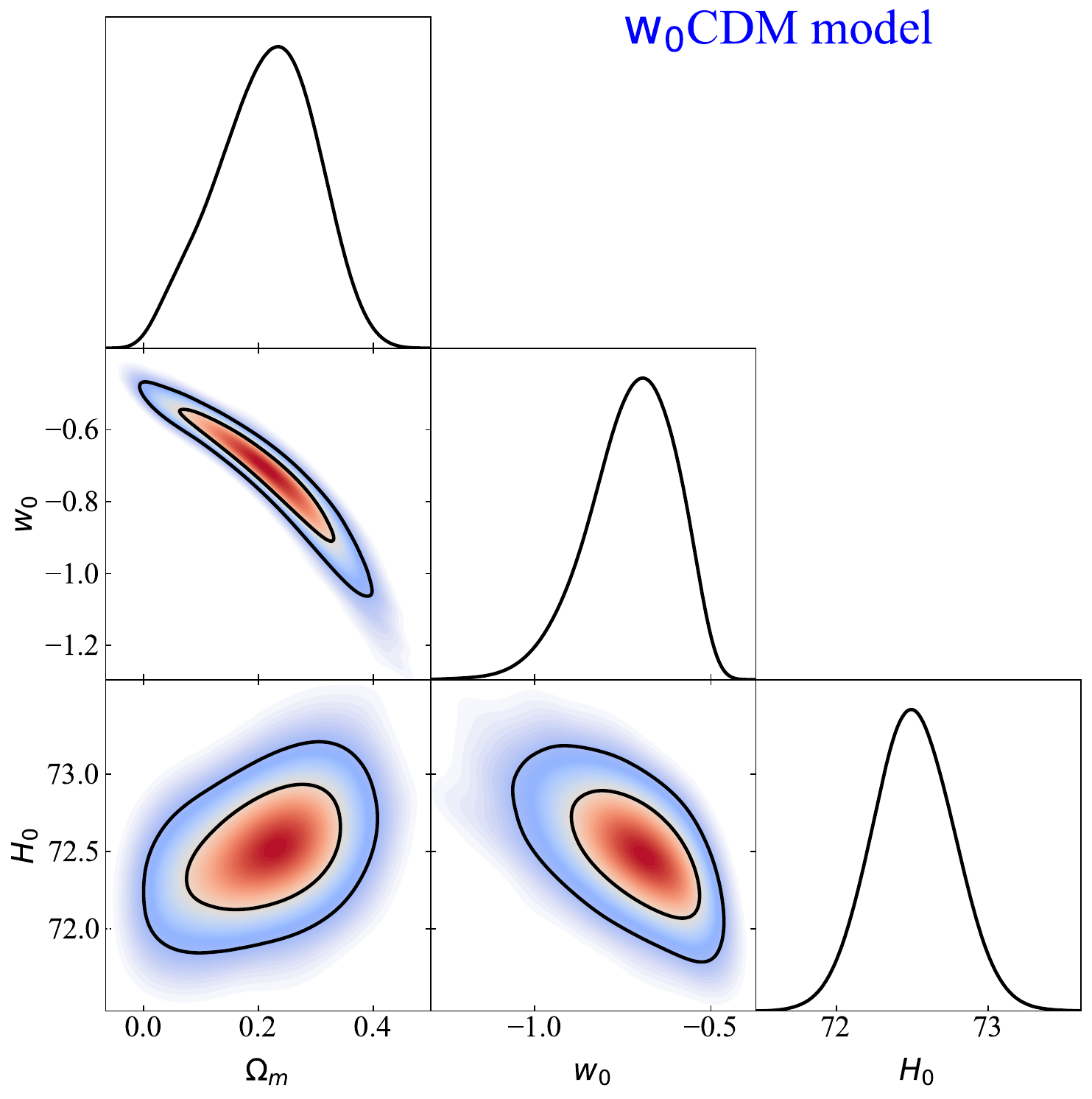}
		\includegraphics[width=0.24\textwidth]{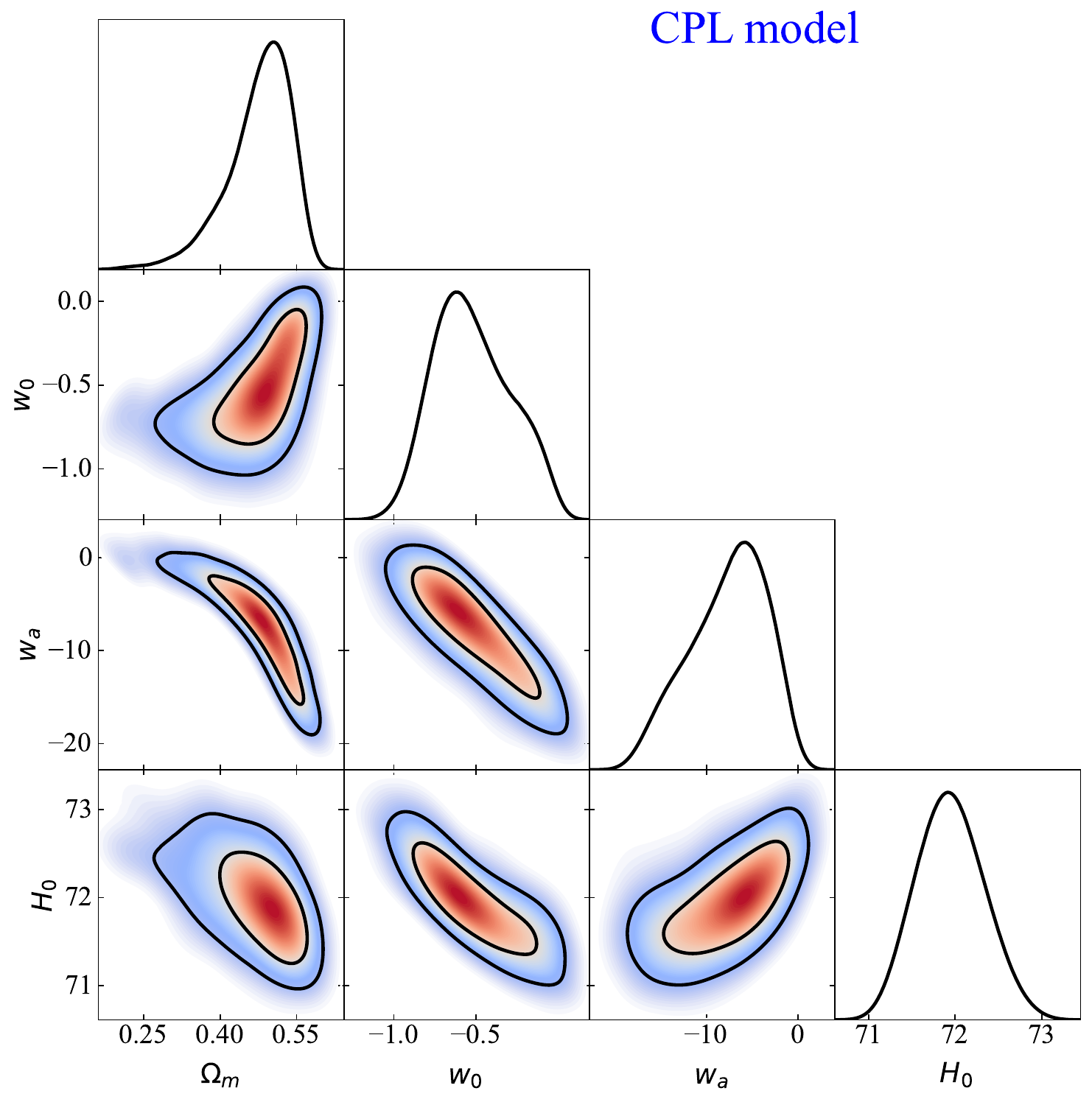}
		\caption{Confidence contours ($1\sigma$ and $2\sigma$) and marginalized likelihood distributions for the parameters space ($\Omega_{m}$, $w_{0}$, $w_{a}$ and $H_{0}$) employing the Pantheon+ sample in the dark energy models. Left and right panels show the results from the $w$CDM and CPL models, respectively.}
		\label{AF2}       
	\end{figure}

	\begin{figure}[h]
		\centering
		\includegraphics[width=0.3\textwidth]{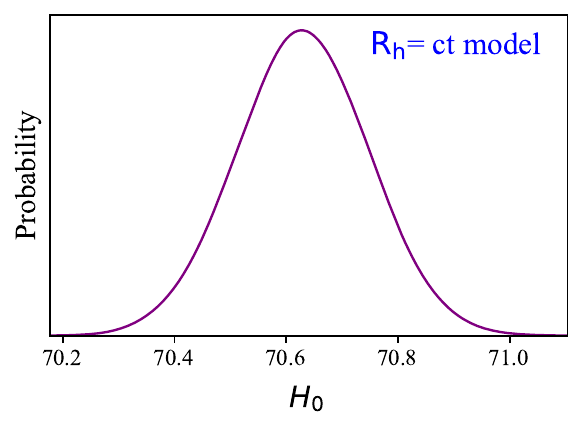}
		\caption{Probability distribution of $H_{0}$ employing the Pantheon+ sample from the $R_{h}$ = ct model.}
		\label{AF3}       
	\end{figure}

	\label{lastpage}
\end{document}